\documentclass[useAMS,usenatbib]{mn2e}

\newcommand{\fma}[1]{\mbox{$#1$}}

\newcommand{\gtsim}{\raisebox{-0.5ex}{$\;\stackrel{>}{\scriptstyle \sim}\;$}}
\newcommand{\unit}[1]{\ifmmode \:\mbox{\rm #1}\else \mbox{#1}\fi}
\newcommand{\mone}{\fma{^{-1}}}

\newcommand{\ha}{H$\alpha$}
\newcommand{\hb}{H$\beta$}
\newcommand{\hy}{H$\gamma$}

\newcommand{\hi}{H~{\sc i}}
\newcommand{\hii}{H~{\sc ii}}
\newcommand{\hei}{He~{\sc i}}

\newcommand{\Oi}{O~{\sc i}}

\newcommand{\nai}{Na~{\sc i}}
\newcommand{\feii}{Fe~{\sc ii}}
\newcommand{\ffeii}{[Fe~{\sc ii}]}

\newcommand{\fex}{[Fe~{\sc x}]}
\newcommand{\fexi}{[Fe~{\sc xi}]}
\newcommand{\tiii}{Ti~{\sc ii}}
\newcommand{\scii}{Sc~{\sc ii}}
\newcommand{\caii}{Ca~{\sc ii}}
\newcommand{\caiii}{Ca~{\sc iii}}
\newcommand{\mgii}{Mg~{\sc ii}}
\newcommand{\Siii}{Si~{\sc ii}}

\newcommand{\kms}{\unit{km~s\mone}}

\newcommand{\gsim}{\!\!\!\phantom{\ge}\smash{\buildrel{}\over
{\lower2.5dd\hbox{$\buildrel{\lower2dd\hbox{$\displaystyle>$}}\over
\sim$}}}\,\,}
\newcommand{\lsim}{\!\!\!\phantom{\le}\smash{\buildrel{}\over
{\lower2.5dd\hbox{$\buildrel{\lower2dd\hbox{$\displaystyle<$}}\over
\sim$}}}\,\,}

\title[Circumstellar envelope ejection in SN 1994W]
{The type IIn supernova 1994W: evidence for the explosive
ejection of a circumstellar envelope
}
\author[Nikolai N. Chugai, Sergei I. Blinnikov, Robert J. Cumming et al.]
{Nikolai N. Chugai$^1$\thanks{E-mail:
nchugai@inasan.rssi.ru (NNC); robert@astro.su.se (RJC); peter@astro.su.se (PL)},
Sergei I. Blinnikov$^2$,
Robert J. Cumming$^3$, Peter Lundqvist$^3$, \newauthor
Angela Bragaglia$^4$,  Alexei V. Filippenko$^5$, Douglas
C. Leonard$^6$, Thomas Matheson$^{5,7}$, \newauthor
Jesper Sollerman$^3$\\
$^1$Institute of Astronomy, RAS, Pyatnitskaya 48, 109017 Moscow,
 Russia\\
$^2$Institute for Theoretical and Experimental Physics, 117218 Moscow, Russia\\
$^3$Stockholm Observatory, Department of Astronomy, Stockholm
University, AlbaNova University Center, SE-106~91 Stockholm, Sweden\\
$^4$Osservatorio Astronomico di Bologna, via Ranzani 1, 40127 Bologna,
Italy\\
$^5$Department of Astronomy, University of California, Berkeley,
CA 94720-3411 USA\\
$^6$Five College Astronomy Department, University of Massachusetts,
Amherst, MA 01003-9305 USA\\
$^7$Harvard-Smithsonian Center for Astrophysics, 60 Garden Street, Cambridge,
MA 02138 USA
}

\begin{document}

\date{
 Accepted 12 May 2004. 
      Received 11 May 2004;
      in original form 29 March 2004.
}

\pagerange{\pageref{firstpage}--\pageref{lastpage}}
\pubyear{2004}

\maketitle

\label{firstpage}
\begin{abstract}

We present and analyse spectra of the Type IIn supernova 1994W
obtained between 18 and 203 days after explosion.  During the luminous
phase (first 100 days) the line profiles are composed of three major
components: (i) narrow P-Cygni lines with the absorption minima at
$-700$ km s$^{-1}$; (ii) broad emission lines with blue velocity at
zero intensity $\sim 4000$ km s$^{-1}$; and (iii) broad, smooth wings 
extending out to at least $\sim 5000$ km s$^{-1}$, most apparent in 
H$\alpha$. These components are identified
with an expanding circumstellar (CS) envelope, shocked cool gas in
the forward post-shock region, and multiple Thomson scattering in the
CS envelope, respectively.  The absence of broad P-Cygni lines from
the supernova (SN) is the result of the formation of an optically
thick, cool, dense shell at the interface of the ejecta and the CS
envelope.  Models of the SN deceleration and Thomson scattering wings
are used to recover the density ($n \approx 10^9$ cm$^{-3}$), radial
extent [$\sim (4-5) \times 10^{15}$ cm] and Thomson optical depth
($\tau_{\rm T}\gsim 2.5$) of the CS envelope during the first month.
The plateau-like SN light curve is reproduced by a hydrodynamical
model and is found to be powered by a combination of  internal
energy leakage after the explosion of an extended pre-supernova ($\sim
10^{15}$ cm) and subsequent luminosity from circumstellar interaction.
The pre-explosion kinematics of the CS envelope is recovered, and is close to
homologous expansion with outer velocity $\sim 1100$ km
s$^{-1}$ and a kinematic age of $\sim 1.5$ yr.  The high mass
($\sim 0.4~M_{\odot}$) and kinetic energy ($\sim 2 \times 10^{48}$
erg) of the CS envelope, combined with small age, strongly suggest that
the CS envelope was explosively ejected $\sim 1.5$ yr prior to the SN
explosion.

\end{abstract}


\section{Introduction} \label{sec-intro}

Recent studies of supernovae (SNe) have made considerable progress in
our knowledge of what makes a star explode.  From observations of
light curves and spectra, theories of pre-supernova evolution are now
better constrained than ever.  Conventional wisdom says that Type II
SNe (SNe~II) are caused by core collapse in massive, usually red
supergiant stars.  In general, theory has no trouble accounting for
the main features of the spectra and light curves of these objects.
However, some SNe~II are remarkably different, and many of their
observational properties are not yet understood.

These events, sometimes known as narrow-line Type II SNe or SNe~IIn
\citep{schlegel90,filippenko97}, show in early spectra the presence of
strong, narrow Balmer emission lines on top of broad emission lines.
Examples of prominent SNe~IIn include 1978K \citep{ryder93}, SNe 1983K
\citep*{niemela85}, 1984E \citep{dopita84}, 1986J
\citep{rupen87,leibundgut91}, 1987F \citep{filippenko89,ws95}, 1988Z
\citep{filippenko91,ss91,turatto93,aretxaga99}, 1994W
(\citealt*{clm94}; \citealt{meikle94}; \citealt*{scl98}, hereafter
SCL98), 1995G \citep{pastorello02} and 1995N \citep{fransson02}. More
recent examples of SNe~II with narrow lines are SNe 1997cy
\citep{germany00,turatto00}, 1998S
\citep{fm98,bowen00,leonard00,fassia01,pozzo04,fransson04}, 1999E
\citep{filippenko00,rigon03} and, most remarkably, SN 2002ic, which
has been identified as a likely SN~Ia in a dense circumstellar (CS)
envelope \citep{hamuy03}, are more recent examples of SNe with narrow
lines.

The general wisdom is that the narrow lines of SNe~IIn originate from the
ionized, dense circumstellar gas \citep{hb87,filippenko91}.  The
interaction of SN ejecta with the dense CS gas modifies the SN optical
spectrum and may fully power the SN~IIn luminosity
\citep{chugai90,chugai92}.  Interaction with the dense CS gas is
indicated also by strong radio and X-ray flux detected in some
SNe~IIn.  SNe~IIn are diverse \citep{filippenko97}, probably
reflecting variations in CS gas density and structure (smooth
vs. clumpy) and SN ejecta parameters (mass and energy).  On top of
this there is also the possibility of asymmetry of both the ejecta and
CS gas. Nevertheless, the primary factors responsible for the
diversity (mass of main-sequence star, explosion mechanism, structure
of pre-supernova, mass-loss mechanism and history, and progenitor
binarity) are still unknown.  All these uncertainties, along with the
possibility of using them to probe pre-supernova behaviour prior to
core collapse, add to the interest in these phenomena.

SN 1994W, discovered on 1994 July 29 (UT dates are used throughout this paper) 
at the pre-maximum phase \citep{cv94}, is among the brightest known SNe~IIn. 
Its proximity (25$\pm$4 Mpc; SCL98) and early discovery made
it an ideal case for detailed study. Preliminary interpretation of the
spectra showed that SN~1994W exploded in the dense CS envelope with a
characteristic radius of $\sim 10^{15}$ cm, while the amount of ejected
$^{56}$Ni was low ($<0.015~M_{\odot}$; SCL98).

Here we further advance our understanding of both the CS
envelope properties and the phenomenon of SN~1994W as a whole by
analysing the spectra and photometry discussed by SCL98 as well as
other, hitherto unpublished, spectra.  In our study we take two
different approaches to modelling the data.  The first is the simulation of the
H$\alpha$ profile, which provides an efficient and straightforward
probe of the Thomson optical depth and density of the CS envelope.
The second approach is a hydrodynamical simulation of the SN explosion
and the light curve based upon the upgraded version of the code {\sc
stella} \citep{blin93j,blin87a}.  An early version of this code proved
effective in computing the hydrodynamics and light curve for
interaction with a dense CS envelope in the case of SN~1979C
\citep{bb93}.

This paper has the  following structure. We start with a description of the
data (Sections \ \ref{sec-obs} and \ref{sec-results}) and the general
picture which arises from a qualitative analysis (Section \
\ref{sec-picture}). Next we quantify parameters of the CS envelope
using a thin-shell deceleration model and a line-profile simulation
(Section \ref{sec-line}).  We then use the gross parameters of
the recovered density distribution as input for detailed
hydrodynamical modelling (Section \ref{sec-hydro}).  Our findings are
summarized and discussed in Section \ref{sec-discussion}.

\section{Observations} \label{sec-obs}

\subsection{Photometry}

The photometric data were compiled and described in SCL98.  In
our light-curve models in Section \ref{sec-results} we use the
photometry up to day 197, assuming SN 1994W exploded on 1994 July
14.0 (SCL98). For a discussion
of the  explosion date, see Section \ref{sec-results}.

\subsection{Spectroscopy}

\begin{table*}

\caption{Log of spectroscopic observations}\label{tab-obs-log}
\vspace{\baselineskip}

\begin{tabular}{@{}llllcccl}
\hline
Date & Phase$^{\rm (a)}$ &Telescope/   & Spectral       & Flux & Seeing   & Wavelength    & Observers \\
  (UT)                &  &spectrograph & res. (\AA) & standard & (arcsec)  & coverage (\AA)&  \\
\hline
1994 07 31.9 & 17.9 & BAO/BFOSC & 13 & BD+25$^\circ$3941 & 1.6 &  4600-7800 & AB \\
1994 07 31.9 & 17.9 & INT/IDS & 0.25 & --- & 2.6--3.0 &   6507--6675 & EZ \\
1994 08 04 & 21.5 & Lick/Kast & 7 &  BD+17$^\circ$4708 & 2.5 &
  4250--7020 & AJB, AF, CP \\
1994 08 13.9 & 30.9 & WHT/ISIS& 2.7  & SP1337+705 & 1.1--1.7 &
  4290--8005 & MB \\
1994 09 01 & 49.5 & Lick/Kast & 5 & BD+28$^\circ$4211,  & 4+ &
  3120--10400 &  AJB, AF, TM \\
                 &  &  &  & BD+26$^\circ$2606 &    &  &  \\
1994 09 08.9 & 56.9 & WHT/ISIS& 11   & Grw+70$^\circ$5824 & 0.9--2.9 &
  3300--9360 & RR \\
1994 09 08.9 & 56.9 & WHT/ISIS& 2.7  & Grw+70$^\circ$5824 & 1.0--1.5 &   4255--5060, &
RR \\
                                    &  & &   & &  &   6345--7150 & \\
1994 09 09 & 57.5 & Lick/Kast & 4 & BD+28$^\circ$4211,
 & 3--5 &   3140--9920 & AJB, AF, TM \\
                 &  &  &  & BD+26$^\circ$2606, &    &  &  \\
                 &  &  &  & BD+17$^\circ$4711 &    &  & \\
1994 10 01 & 79.5 & Lick/Kast & 4 & Feige 110,        & 2--3 &
  3130--8040 & AF, LH, TM \\
                 &  &  &  & BD+26$^\circ$2606 &    &  & \\
1994 10 11 & 89.5 & Lick/Kast & 4 & Feige 110,        & 3--4 &
  3160--8020 & AF, TM \\
                 &  &  &  & BD+26$^\circ$2606 &    &  & \\
1994 11 12.3 & 121.3 & INT/IDS & 6    & --- & 1.6   & 5485-7370 & RR, NO \\
1995 01 26.6 & 196.6 & Keck/LRIS & 6 &  HD 19445 & 1.2   & 5628--6939 &
AF, LH \\
1995 02 01.9 & 202.9 & NOT/LDS & 12   & Feige~34 & 0.9   & 5000-10000 &
JS, RC \\
\hline
\end{tabular}

{\footnotesize \raggedright
Note: (a) Days after 1994 July 14.0. (b)
Observers: AB -- A. Bragaglia; AF -- A. Filippenko; AJB -- A. J. Barth; CP --
C. Peng; EZ -- E. Zuiderwijk; JS -- J. Sollerman; LH -- L. Ho; MB -- M. Breare;
NO -- N. O'Mahony; TM -- T. Matheson; RC -- R. Cumming; RR -- R. Rutten.}
\end{table*}

  \begin{figure*}
  \centering
  \vspace{14.5cm}
  \includegraphics{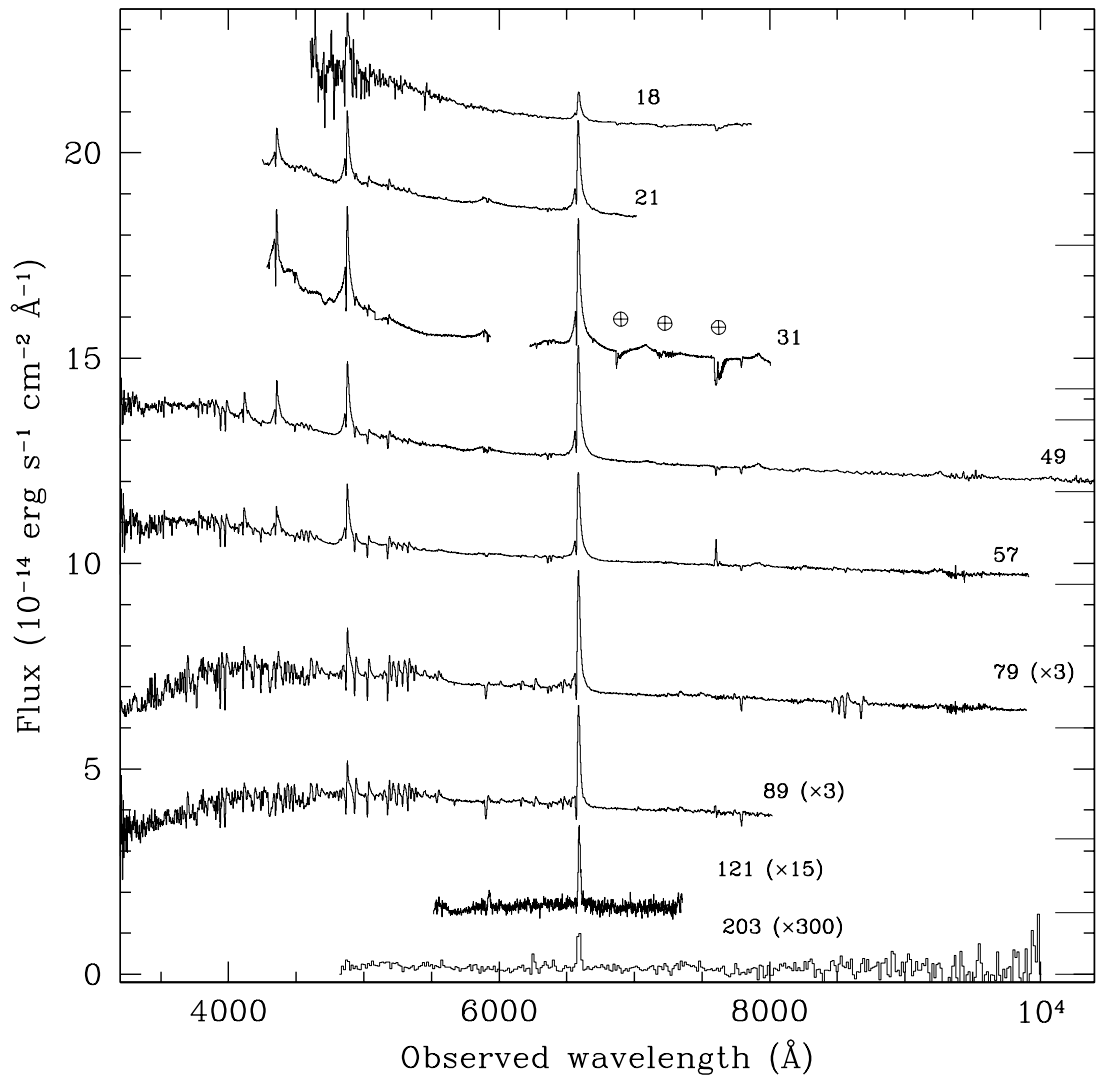}
  \caption[Spectra of SN 1994W: days 18--203]{Spectra of SN 1994W: days
18--203.  The spectra have been shifted vertically for clarity.
No heliocentric correction has been applied to the wavelength scale.  The
ticks on the right-hand side mark the zero level for each spectrum.  The
spectra from day 79 and later have been multiplied by a constant,
noted in parentheses.  The La Palma spectra from days days 18 and 57 and the day 197 Keck
spectrum are not shown. Telluric absorption lines in the day 31 spectrum are
marked.
}
  \label{f-spectra}
  \end{figure*}

  \begin{figure*}
  \centering
  \vspace{14.5cm}
  \includegraphics{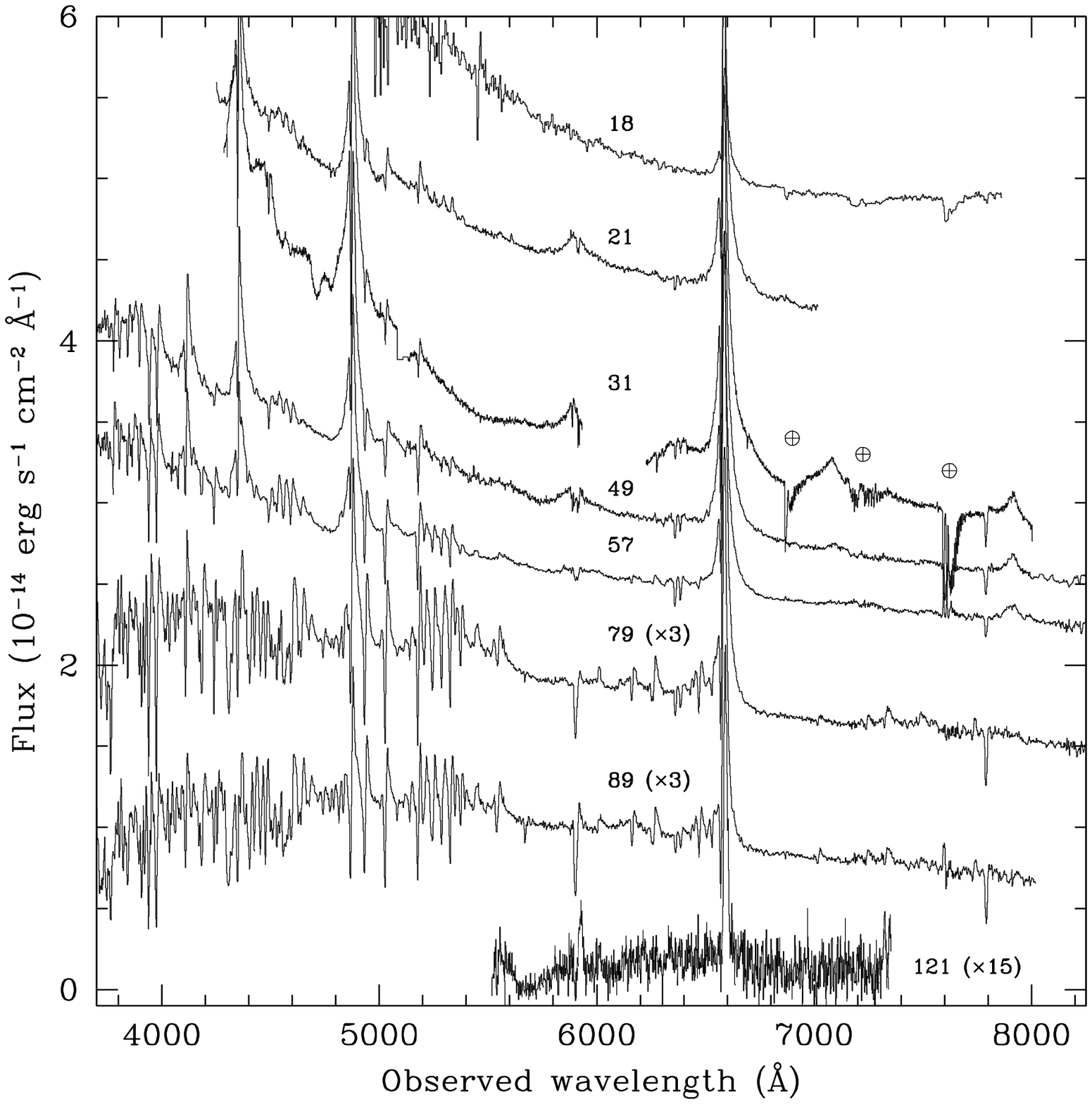}
  \caption[Spectra of SN 1994W: days 18--121]{Spectra of SN 1994W: days
18--121.  Same as Fig.\  \ref{f-spectra}, but with the ordinate
stretched to show the weaker features more clearly. The day 197 and 203 spectra
are omitted. Telluric absorption lines in the day 31 spectrum are marked.
}
  \label{f-spectra-dyn}
  \end{figure*}

The spectroscopic observations are described in Table \ref{tab-obs-log}.
Some of these, taken from La Palma, were presented in Table 2 of SCL98.
We supplement them with spectra taken at the Lick, Keck,
and Bologna Observatories. The data cover epochs 18--203 days
after explosion.  We now describe these spectra in detail.

The La Palma spectra were reduced using the {\sc figaro}
\citep{shortridge90} and NOAO {\sc iraf}\footnote{IRAF (Image Reduction and 
Analysis Facility) is distributed by the National Optical Astronomy 
Observatories, which are operated by the Association of Universities for 
Research in Astronomy, Inc., under cooperative agreement with the US 
National Science Foundation.} packages.  The CCD frames were
bias-subtracted, flat-fielded, and corrected for distortion and cosmic rays.
Wavelength calibration was carried out using arc spectra of copper-neon and
copper-argon lamps, and should be accurate to better than 0.1~\AA\ for the
high-resolution WHT and INT spectra, 0.5~\AA\ for the low-resolution WHT and INT
spectra, and 1~\AA\ for the NOT spectrum.

The Bologna  spectrum was obtained with the Bologna Astronomical
Observatory 1.5-m telescope with the spectrograph BFOSC on 1994 July
31.9 and reported by \citet{bmb94}.

Spectra were also obtained at the Shane 3~m reflector at
Lick Observatory on days 21, 49, 57, 79 and 89, and at the W. M. Keck
II 10~m telescope on day 197.  The Lick observations were made with
the Kast double spectrograph \citep{ms93}.  The Keck spectrum was
obtained using the Low Resolution Imaging Spectrometer (LRIS;
\citep{oke95}) with a 1200 lines~mm$^{-1}$ grating.  For details, see
Table \ref{tab-obs-log}.

For the Lick data, all one-dimensional sky-subtracted spectra were
extracted in the usual manner, and each spectrum was then wavelength and
flux calibrated, as well as corrected for continuum atmospheric extinction
and telluric absorption bands \citep{wh88,bessell99,matheson00}.

The spectra were flux-calibrated by comparison with the standard stars
listed in Table \ref{tab-obs-log}.  The sky position of SN 1994W in
late 1994 meant that the observations were made when the supernova was
at rather low altitude, and correspondingly high airmass.
We corrected for the difference in atmospheric extinction brought
about by the differing zenith distances of the supernova and the
standard star.  Airmasses at different zenith distances were taken from the
relation given by \citet{murray83}, and we used the tables of extinction
coefficient versus wavelength for La Palma given by \citet{king85}.

With two exceptions, all our spectra were taken at the
parallactic angle \citep{filippenko82}, minimising slit losses due
to atmospheric dispersion.
The first exception was the day 18 Bologna spectrum of SN 1994W, which
shows considerable loss of flux in the blue.  We corrected the slope
of the spectrum to match the $B$ and $V$ photometry from the same
night reported by \citet{bmb94} using filter functions from
\citet{gm01}.

The second exception was the standard-star observation for the day 31 La
Palma spectrum.  While the SN 1994W spectrum was taken at the
parallactic angle, the standard-star spectrum was taken at an airmass of
1.6, with the slit at position angle 33$^\circ$ when the parallactic
angle was 119$^\circ$.  No contemporary filter photometry is available
for correcting the supernova spectrum.  The effect on the day 31
supernova spectrum depends on how the standard star was positioned in
the slit, which in turn depends on the combination of the spectral
response of the acquisition camera at the telescope.  The acquisition
camera at the WHT in 1994 August was a Westinghouse ETV-1625 whose
response curve peaked at 4400~\AA, with half-power points at 3400~\AA\
and 6200~\AA\ (C. Jackman \& D. Lennon, 2004, private communication).  The
standard star, Grw +70$^\circ$5824, is a DA3 white dwarf with a blue
spectrum.  It therefore seems likely that the slit was positioned so
that maximum transmission was at around 4400~\AA, at the blue end of
the spectrum.

We have estimated the amount of flux that would be lost with
wavelength, and find that the losses would largely have been at the
red end of the standard spectrum --- as much as 80 per cent of the flux
at 8000~\AA.  The effect on the calibrated supernova spectrum was thus
to make it appear redder than it really was.  Since the
day 31 spectrum is already remarkably blue (Section \ref{sec-d31}), we
have made no further correction to the spectrum; the calibration can
be regarded as a conservative estimate of the already extreme
continuum slope.

No observations of flux standards were available for the INT spectra
taken on days 18 and 121.  A very rough flux-calibration of the day 18
INT spectrum was made by assuming that the flux in the wings of the
\ha\ line (at the red and blue edges of the spectrum) declined at the
same rate as the visual magnitude estimates between days 18 and 31,
i.e., by a factor of about 1.6 in $f_\lambda$.  A similarly rough
calibration of the day 121 spectrum was made by comparing with our $R$-band
photometry from day 123 (SCL98).

The spectrum taken on day 197 was scaled so that the
flux in the narrow \ha\ feature matched that in the day 203 spectrum.
The continua then match well in the overlap region.

Spectra from days 18, 57, 121 and 203 were displayed in SCL98 (their
Figs. 2 and 3). They also showed the portion of the spectrum on day 31
covering \ha.

As a final correction, we noted that all
the spectra cover the entire $V$ band, and we
scaled each spectrum to match contemporary $V$-band photometry.
The procedure was as follows.  We assumed that the day 57 La Palma
spectrum was correct. Next, we interpolated the $V$ magnitudes in the
light curve presented by SCL98.  This gave us relative fluxes in $V$
for all the epochs.  We then integrated all the plateau spectra (days
21--89), multiplying them by a standard $V$ filter function \citep{allen73}
with transmission values interpolated between points at 100-\AA\
intervals.  The whole spectrum was then scaled linearly in flux so
that the integrated $V$ magnitude corresponded to the $V$-band
photometry in SCL98.

\section{Results} \label{sec-results}

\begin{figure*}
  \centering
  \vspace{11cm}
  \includegraphics{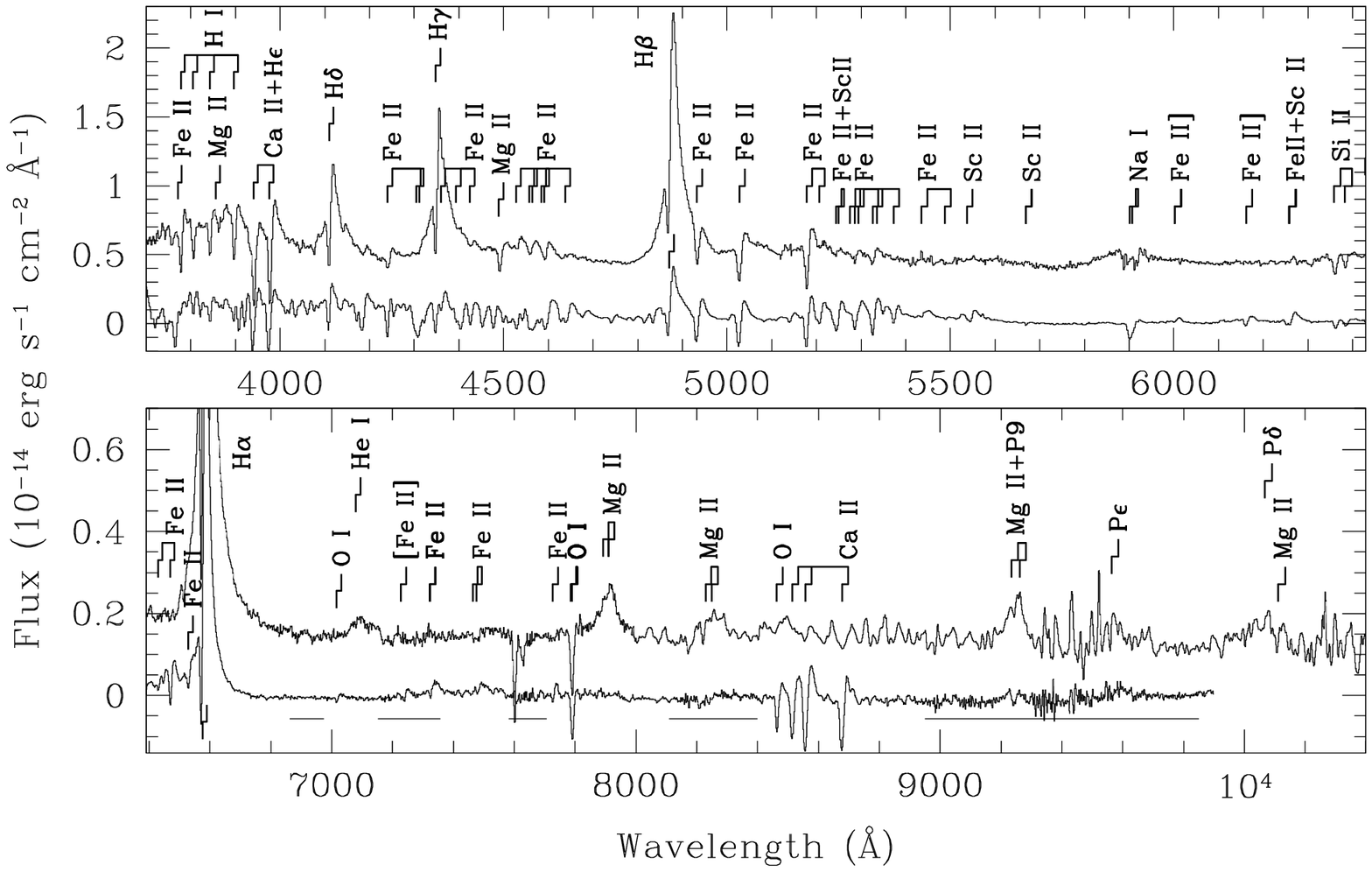}
  \caption[Line identifications.]{Line identifications for days 49
(upper spectrum) and 79.  The spectra have been continuum-subtracted and
shifted for clarity.  Each line is indicated by a broken vertical bar
marking heliocentric velocities $-$700 \kms\ and 0 \kms.  Horizontal bars longward of 6700~\AA\ mark
regions where telluric absorption has been removed.}
  \label{f-idents1}
  \end{figure*}

During the period of our observations, the spectrum of SN 1994W
(Figs. \ref{f-spectra} and  \ref{f-spectra-dyn}) was dominated by narrow Balmer emission lines,
initially with P-Cygni absorption features, and accompanied by a wide
variety of weaker lines showing broad emission and/or narrow P-Cygni
profiles \citep[cf.][]{fb94}. Line identifications are presented in
Table \ref{tab-line-ids1} and in Fig.\ \ref{f-idents1}.  In the
following sections, we examine the spectral development for each epoch
in turn, and then concentrate on the individual features and ionic
species.

In this section, quoted line fluxes (see Table \ref{tab-lines1}) were
measured by interpolating the neighbouring continuum across the line
in question.  The line fluxes are net fluxes and do not correct for
the narrow P-Cygni absorption components.


  \begin{figure}
  \centering
  \vspace{8.5cm}
  \includegraphics{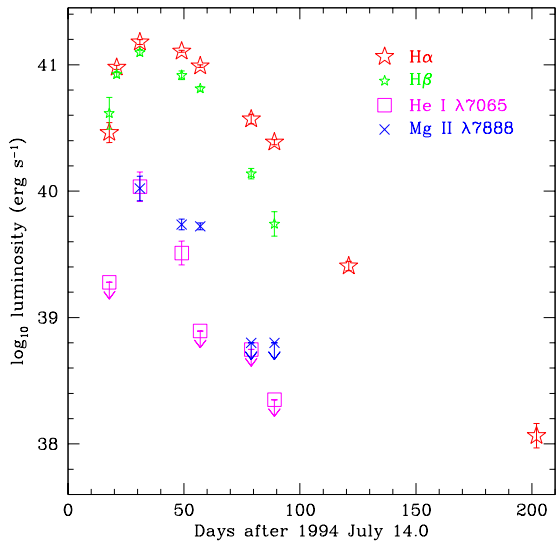}
  \caption[Line luminosities]{Luminosity evolution of various lines.
  Line fluxes are detailed in Table \ref{tab-lines1}. A distance of
  25.4 Mpc and an extinction of $E_{B-V} = 0.17$ mag were adopted. }
  \label{f-luminosities}
  \end{figure}

\begin{table*}

\caption{Line identifications for days 18--121.  Symbols are as follows: a --- line seen in
absorption only; b --- broad emission, e --- emission; p --- P-Cygni
profile; -- (dash) --- no feature detected, : (colon) ---
uncertain identification.  The \feii\ lines were identified with the help
of the line lists of \citet*{fuhr88} and \citet{sp03}.
Wavelengths in \AA\ are from these references and from \citet{vanhoof99}.}
\label{tab-line-ids1}
\vspace{\baselineskip}
\begin{tabular}{lcccccccclcccccccc}
Identification &18&21&31&49&57&79&89&121&Identification &18&21&31&49&57&79&89&121\\
\hline

\feii\  3764.11 &&&&&&&&&	                               \scii\  5657.90 &&&&&&&\\
+H11 3770.63 & & & &p&p&--&--& & 			      +\scii\ 5658.36 &--&--&--&--&--&a&a& \\
H10  3797.90   & & & &p&p&p&p& &			       \hei\  5875.61 &--&b&pb&b:&b:&-- &--&--\\
H9  3835.38   & & & &p&p&p&p& & 			       \nai\  5889.95  &&&&&&&\\
\mgii\   3849.3   & & & &p&p&--&--& &			      +\nai\  5895.92 &--&b:&b:&ba:&ba:&p&p&ep:\\
H8   3889.05   & & & &p&p&p&p& &			       \feii]   5990.60 &&&&&&&\\
\feii\   3914.51 & & & &--&--&p&p& &			      +\feii] 5991.38	&--&--&--&--&--&ep &ep &--\\
\caii\   3933.66 && & &p&p&p&p& &			       \feii]   6150.10 &--&--&--&--&p&p&p&--\\
\caii\  3968.47  &&&&&&&&& 			               \scii\  6245.63  &&&&&&&\\
+ H$\epsilon$ 3970.07	& && &pb&pb&pb&pb& &		      +\feii\ 6247.58 &--&--&--&pb &pb &pb &pb &--\\
H$\delta$   4101.73 && & &pb&pb&pb&pb& &	               \Siii\  6347.11 &--&p&p&p&p&p&p&--\\
\feii\   4233.12 &--&--&--&p&p&p&p& &			       \Siii\  6371.37 &--&p&p&p&p&p&p&--\\
\feii\   4296.57 &&&&&&&&&		                       \feii\  6416.91 &--&--&--&--&p&p&p&--\\
+\feii\ 4303.17 &--&-- &-- &--&p&p&p& &			       \feii\  6456.38 &--&--&--&--&p&p&p&--\\
H$\gamma$  4340.46 &-- &pb&pb&pb&pb&pb&pb& &		       \feii\  6517.02 &--&--&--&--&p&p&p&--\\
\feii\   4351.76 &--&--&--&--&p&p&p& &			       H$\alpha$  6562.80 &pb&pb&pb&pb&pb&pb&pb&e \\
\feii\   4385.38 &--&--&--&p&p&--&--& &		               \hei\  6678.15 &--&p:&pb&--&--&--&--&--\\
\feii\   4416.83 &--&--&--&p&p&p&p& &			       \Oi\   7002.20 &--&--&--&--&--&p&p&--\\
\hei\    4471.5   &--&--&ba&-- &--&--& & &		       \hei\   7065.22 &--& &b&b&--&--&--&--\\
\mgii\   4481.2   &--&p&p&p&p&--&--& &			       \ffeii\ 7214.71 &--& &--&--&--&p&p&--\\
\feii\   4520.21 &--&--&--&p&p&--&--& &		               \caii] 7291.47  &--& &--&--&--&--&--&e\\
\feii\   4549.46 &&&&&&&&&			               \feii\  7308.07  &&&&&&&\\
+\feii\   4555.88 &--&p&p&p&p&p&p& &			      +\feii\  7310.22 &--& &--&--&--&p&p&--\\
\feii\   4576.34 &&&&&&&&&			               \caii] 7323.89  &--& &--&--&--&--&--&e\\
+\feii\   4582.83 &--&p&p&p&p&p&p& &			       \feii\  7449.34 &--& &--&--&--&p&p& \\
\feii\   4629.34 &--&p&--&p&p&p&p& &			       \feii\  7462.41 &--& &--&--&--&p&p& \\
H$\beta$ 4861.32 &pb&pb&pb&pb&pb&pb&pb& &		       \feii\  7711.72 & & &--&p&p&p&p& \\
\feii\   4923.94 &p: &pb&pb&pb&pb&pb&pb& &		       \Oi\   7773.8    &p& &p&p&p&p&p& \\
\feii\   5018.44 &p: &pb&pb&pb&pb&pb&pb& &		       \mgii\  7877-96  & & &b&b&b&--&--& \\
\feii\   5169.05 &-- &p&p&p&p&p&p& &			       \mgii\  8213.98  &&&&&&&\\
\feii\   5197.59 &-- &p& &p&p&p&p& &			      +\mgii\   8234.64 & & & &b:&b:&--& & \\
\feii\   5234.63 &&&&&&&&&			               \Oi\   8446.36  & & & &p&p&p& & \\
+\scii\ 5239.81 	&--&--&--&--&p&p&p& &		       \caii\   8498.03 & & & &--&--&p& & \\
\feii\   5264-84 &--&p&--&p&p&p&p& &			       \caii\   8542.09 & & & &p&p&p& & \\
\feii\  5316.62 &&&&&&&&&			               \caii\   8662.14   & & & &--&p&p& & \\
+\feii\ 5325.54 	&--&p&p&p&p&p&p& &		       \mgii\ 9218.25 &&&&&&&\\
\feii\  5362.85 &--&--&--&--&p&p&p& &			      +Pa9 9229.01  &&&&&&&\\
\feii\  5425.26 &--&--&--&--&pe&pe&pe& &	              +\mgii\  9244.26 & & & &b&b&b&\\
\feii\  5477.66 &--&--&--&--&--&p&p& &  		       Pa$\epsilon$ 9545.97  & & & &pb:&pb:&pb:&\\
 \scii\  5526.79 &--&--&--&--&pb&pb&pb&p:&  	               Pa$\delta$ 10049.4  &&&&&&&\\
 &&&&&&&&&						      +\mgii\ 10092.1 & & & &b:& & &\\

\hline
\end{tabular}

\end{table*}


\subsection{Spectral evolution}

\subsubsection{Day 18}

The low-resolution Bologna spectrum showed a blue continuum with
prominent lines of \ha\ and \hb, which are narrow with broad wings. The
latter have been interpreted as an effect of multiple Thomson
scattering of line photons in the CS envelope \citep{chugai01}.  Our
high-resolution spectrum for this epoch (Fig.\ 3 in SCL98) covers only
the \ha\ line, but shows a narrow P-Cygni absorption at about $-$800
\kms\ not visible in the low-resolution spectrum.  The emission peak
is close to the velocity of nearby \hii\ emission, measured at
1249$\pm$3 \kms, which we adopt as our best estimate of the
heliocentric velocity of the supernova.  The interstellar \nai\ D
lines give a somewhat lower value, 1185$\pm$15 \kms, measured from the
day 31 spectrum.

\subsubsection{Day 21}

The rapid brightening of the supernova at this epoch (SCL98) was
reflected in increasingly strong Balmer lines.  The spectrum showed a
strong blue continuum, and prominent lines of \ha,
\hb\ and \hy, all exhibiting narrow P-Cygni profiles with broad emission
wings.  The Balmer decrement was remarkably shallow, with flux ratio
\ha:\hb:\hy\ = 1.1:1:0.5 (dereddened assuming $E_{B-V} = 0.17$ mag from SCL98 and
the reddening law of \citealt*{cardelli89}).  The Balmer lines were
accompanied by an asymmetric triangular broad emission feature
corresponding to \hei\ $\lambda5876$, possibly blended with the \nai~D
doublet, and many narrow P-Cygni lines of \feii.

\subsubsection{Day 31}\label{sec-d31}

\begin{figure}
  \vspace{7.4cm}
  \includegraphics{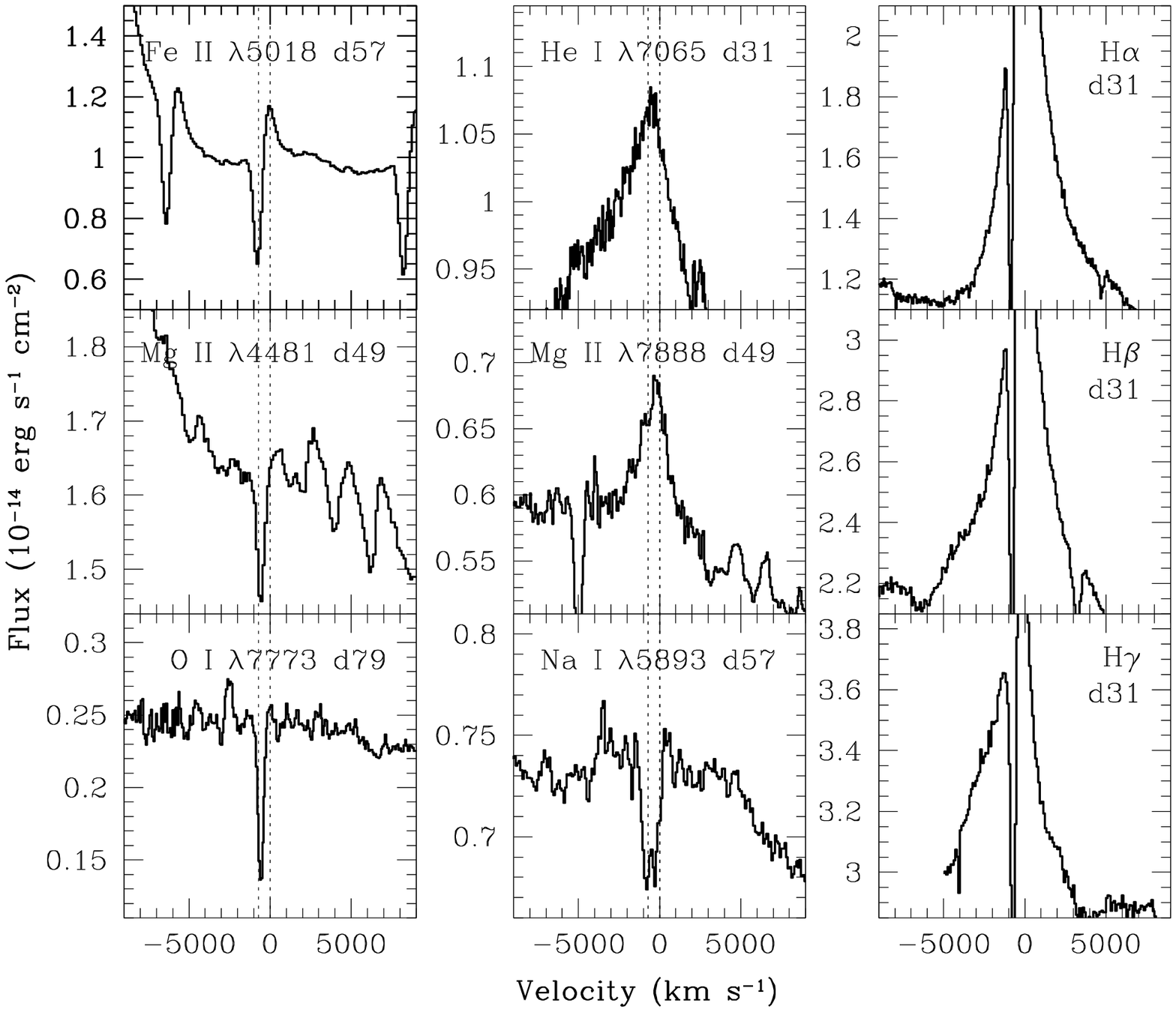}
  \caption[Line profiles.]{Selected line profiles; day numbers are indicated. The left-hand
panel shows examples of narrow P-Cygni profiles, the middle panel
shows broad lines (note the extended wings in \nai\ on day 57), and
the right-hand panel shows the wings of the \hi\ lines on day 31.  The dotted
vertical lines mark velocities of 0 and $-$700 \kms.
}
  \label{f-profiletypes}
  \end{figure}

By day 31, the blue continuum appeared to have steepened.  The
calibration of the spectrum is uncertain, as noted in Section \ref{sec-obs}
above, but the evidence suggests that the spectrum
really was as remarkably blue as it appears to be. The Balmer lines
still show a flat decrement, as on day 21, with measured ratios
1.2:1:0.6.  The blue wings in \hb\ and \hy\ appear to be stronger
relative to the red wings, compared to day 21 (Fig.\
\ref{f-profiletypes}, right-hand panel).

Broad, triangular \hei\ emission lines with full-width at half-maximum velocity
width ($v_{\rm FWHM}$) around 2500 \kms\ were stronger than on day 21 (Figs.\
\ref{f-spectra-dyn} and \ref{f-profiletypes}).  \hei\ $\lambda5876$ was accompanied 
by strong \hei\ $\lambda7065$.  Broad \hei\ $\lambda6678$ appeared to fill out
the red wing of \ha, and the triplet line at $\lambda4471$ is present
as broad emission which underlies neighbouring \feii\ lines.  All but
\hei\ $\lambda7065$ also showed weak, narrow P-Cygni features with
minima around $-$700 \kms.

A similarly triangular emission feature is seen at 7890~\AA, which we
identify with the \mgii\ triplet at 7877--7896~\AA\ (see Section
\ref{sec-metals}).

Stronger narrow P-Cygni lines with absorptions around $-$600 \kms\ are
seen in \Oi\ $\lambda7773$, \Siii\ $\lambda\lambda6347$, 6371, and a
large number of \feii\ lines.  The \Oi\ and \Siii\ lines are
dominated by absorption, while in \feii\ the equivalent widths of the
absorption and emission components are comparable.

\subsubsection{Day 49}

The day 49 spectrum (Figs.\ \ref{f-spectra}, \ref{f-spectra-dyn} and
\ref{f-idents1}) shows a redder continuum.  The Balmer decrement
steepened to 1.5:1:0.6.  The \hei\ $\lambda5876$ and \nai~D blend was
again asymmetric as on day 21, and appeared to have a broader blue
wing.  The only other remaining \hei\ line, $\lambda7065$, is no
longer as prominent as on day 31.  The \mgii\ $\lambda$7877--7896
feature was still clearly visible.  Similar broad features were
detected at $\sim$8230~\AA\ and $\sim$9220~\AA, lending support to the
identification with \mgii.

Many new \feii\ lines are apparent, plus \caii\ H \& K and possibly the
\tiii\ multiplet at around 3500~\AA.  The wider wavelength coverage picked
up a large number of overlapping narrow P-Cygni lines in the blue;
most of the identified ones are \hi\ and \feii.

\subsubsection{Day 57}

Eight days later, the Balmer decrement showed no significant change at
1.5:1:0.6.  The spectrum showed ever stronger \feii\ lines.  No
unambiguous trace of \hei\ remained (Fig. \ref{f-spectra-dyn}),
suggesting that the emission feature at $\sim$6000~\AA\ was now only
due to \nai~D.  The emission in this feature broadened and flattened
out, extending from $-$4000 \kms\ to +6000 \kms\ (Figs.
\ref{f-spectra-dyn} and \ref{f-profiletypes}).  The \mgii\
$\lambda$7877--7896 feature was still present, but appeared more
flat-topped than on day 49.  P-Cygni lines of \Oi\ $\lambda$8446 and
the \caii\ near-infrared triplet, also weakly present on day 49, were
clearly detected in the red.

\subsubsection{Day 79}

The continuum appeared again to have reddened (Fig.\ \ref{f-spectra}).
The Balmer lines decreased in strength relative to the now very
numerous \feii\ P-Cygni lines (Figs.\ \ref{f-spectra-dyn} and
\ref{f-idents1}).  The \ha/\hb\ ratio was now $2.7 \pm 0.3$, close to
Case~B.  Narrow overlapping absorption and P-Cygni features completely
dominated in the blue.  In the red, the feature at 7890~\AA\ had
disappeared, but \Oi\ $\lambda$8446 and the \caii\ triplet had
increased in strength.  The narrow absorption in \nai~D had increased
markedly in strength since day 57, and broad \nai\ emission was no
longer apparent.

Narrow P-Cygni profiles of \Oi\ $\lambda$7002 and Sc~{\sc ii}
appeared, the latter possibly accompanied by a broad emission
component.  Semi-forbidden lines of \feii] $\lambda\lambda$5991,
6150 were clearly detected for the first time.

\subsubsection{Day 89}

The spectrum on day 89 was essentially the same as on day 79, but was
weaker and showed a redder continuum.  The \ha/\hb\ ratio increased to
$5.2 \pm 1.3$, and \hy\ was now blended with \feii\ lines.  Both \nai~D and \Oi\
$\lambda$7002 continued to increase in strength.

\subsubsection{Day 121}

Our spectrum for day 121 (Figs.\ \ref{f-spectra} and
\ref{f-spectra-dyn}) has smaller wavelength coverage and rather poor
signal-to-noise ratio, but it is obvious that a dramatic change had
occurred since day 89, paralleling the supernova's precipitous fading
after day $\sim$110 (SCL98).

As reported by SCL98, three narrow emission lines are unambiguously
detected: \ha\ with a $v_{\rm FWHM}$ of 780 \kms, plus much fainter \nai~D
($v_{\rm FWHM} = 775 \kms$) with a P-Cygni absorption component, and
\caii] $\lambda$7291 ($v_{\rm FWHM} = 475 \kms$) close
to the red edge of the spectrum.
The emission peak of \ha\ had moved
farther to the red than on day 89, at about +70$\pm$100 \kms.  The
\scii\ P-Cygni feature at 5530~\AA\ appears also to have survived since
day 89, showing a sharp absorption feature at $-800 \pm 100$ \kms.

Underlying the narrow features is what appears to be a low, undulating
continuum.  SCL98 argued that the sharp drop to zero flux at
5620~\AA\ was evidence that the continuum was rather composed of
broad, overlapping emission lines.  The steepness of the drop at 5620~\AA\
indicates that the velocity widths of such lines can be no
greater than $\sim$5000 \kms.  Similar features have been seen in late-time
spectra of the SN~IIn 1997cy \citep{germany00,turatto00}.  A sharp
drop at this wavelength is in fact also typical of SNe~Ia and some SNe~Ib/Ic
at late times, when the emission is dominated by lines of \feii\
and \ffeii\ \citep[see][]{axelrod80}.  These similarities suggest that SN
1994W had reached its nebular phase at this epoch.

\subsubsection{Day 197}

Our spectrum, not shown in Figs.\ \ref{f-spectra} and
\ref{f-spectra-dyn}, detected only \ha, which was perhaps just
resolved.  Its velocity width was $v_{\rm FWHM} \approx 625$ \kms,
consistent with no change since day 121.  \nai\ emission at the same
strength relative to \ha\ as on day 121 would not have been detected, 
given the low signal-to-noise ratio of the spectrum.

\subsubsection{Day 203}

The day 203 spectrum (Fig.\ \ref{f-spectra}) confirms that narrow,
unresolved \ha\ is the only detectable feature from the supernova.
The galaxy background probably dominates the remaining, noisy
continuum, though photometry indicates that the supernova can account
for up to 50 per cent of the flux (SCL98).  SCL98 put limits on the
flux in broad lines of width $\sim$4500 \kms.

\subsection{General spectral appearance}

\subsubsection{Line profiles}

The spectra reveal two major types of line
profile-forming components (SCL98; Fig. \ref{f-profiletypes}): narrow P Cygni
profiles and broad lines.  However, careful inspection reveals that
while the blue velocity at zero intensity (BVZI) 
of the broad lines does not exceed $-5000$ km s$^{-1}$, \ha\ at
the luminous stage shows a red wing extending at least up to
+7000 km s$^{-1}$.  This wing is clearly detectable in \ha\ on
days 31, 49 and 57 (Figs.\  \ref{f-spectra-dyn} and
\ref{f-profiletypes}).  We believe that this wing is caused by a
Thomson scattering effect in the expanding CS envelope (see also
Section \ref{sec-picture}).  Since the intensity of the
Thomson scattering wing is proportional to the line intensity, the
wing may be present but not obvious in other, weaker lines.

The intrinsically broad lines typically have a triangular profile.  On
day 31 they show an apparent skewing toward the blue, possibly due to
an occultation effect.  The \nai\ doublet on day 57 shows a
flat-topped profile with BVZI$\approx 4000$ km s$^{-1}$.  No other
line shows such a flat-topped profile, nor does \nai\ doublet at other
epochs, though the feature is dominated by \hei\ $\lambda5876$ earlier
than this.  This suggests that the material emitting \nai\ has a
narrower distribution of velocity than the other broad lines (see
Section \ref{sec-broad}).

\subsubsection{Line strengths}

Assuming a distance of 25.4 Mpc to NGC 4041 and $E_{B-V} = 0.17$ mag (SCL98), we
have plotted the luminosity evolution of the strongest emission lines in
Fig.\  \ref{f-luminosities}.  Note the quasi-exponential decline of the
H$\alpha$ luminosity with exponential lifetime $\sim 21$ days. This fast
decline is in dramatic contrast to other SNe~IIn, for example SN~1995G, whose
narrow H$\alpha$ luminosity dropped by only a factor of three during the
first 600 days \citep{pastorello02}.  The fast decay of the H$\alpha$
luminosity is consistent with a small amount of $^{56}$Ni in SN~1994W (SCL98)
and also indicates the absence of dense CS gas at large distances from the
pre-supernova (see Section \ref{sec-discussion}).

\begin{table*}
\centering
\begin{minipage}{150mm}

\caption{Flux measurements  of emission lines, in units of 10$^-14$ erg s$^{-1}$ cm$^{-2}$.}\label{tab-lines1}
\vspace{\baselineskip}

\begin{tabular}{@{}lccccccccc}
  & \multicolumn{9}{c}{Day} \\
Line   & 18 & 21 & 31 & 49 & 57 & 79 & 89 & 121 & 203 \\
\hline
\hy\  4340.46 & --- & 30(2) & 47(4) & 34(1.5) & 24.5(2)  & --- & --- & --- & --- \\
\hb\  4861.32 & 30(10) & 61(3) & 91(5) & 60(5) & 47(2) & 10(1) & 4(1) & --- & --- \\
\feii\  5018.44 & --- & 1.7(0.1) & 0.7(0.1) & 2.1(0.5) & 2.7(0.7) & 2.0(0.7) & 1.6(0.2) & --- & --- \\
\hei\  5875.62 &  &  &  &  &  &  &  &  &  \\
+\nai$^{\rm (a)}$   5889-95 & --- & 13.6(2) & 14(4) & 12(1) & 5(1) & $<$1 & $<$1 & --- & --- \\
\nai$^{\rm (b)}$ 5889-95 & --- & --- & --- & --- & 0.14(0.05) & 0.2(0.02) & 0.6(0.1) & 0.11(0.03) & $<$0.03 \\
\ha\  6562.80 & 25(5) & 82(3) & 130(6) & 110(2) & 84(2) & 32(3) & 21(1) & 2.2(0.2) & 0.1(0.025) \\
\hei\  7065.22 & $<$2 & --- & 9.7(3) & 2.9(0.7) & $<$0.7 & $<$0.5 & $<$0.2 & --- & --- \\
\mgii$^{\rm (a)}$  7877-7896 & --- & --- & 10(2.5) & 5.2(0.5) & 5.0(0.3) & $<$0.6 & $<$0.6 & --- & --- \\
\mgii\  9218-44 & --- & --- & --- & --- & 6.3(1) & 7.5(0.8) & 1.0(0.3) & --- & --- \\
\hline
\end{tabular}

{\footnotesize \raggedright Notes: (a) blend, (b) emission
component only.\newline \noindent  Error estimates are 1$\sigma$.  Upper limits are 3$\sigma$. }

\end{minipage}
\end{table*}

\subsubsection{Continuum}\label{sec-continuum}

We carried out black-body fits to the continuum.  We obtain colour
temperatures of $13000 \pm 2000$ K on day 21, 15000 K on day 31, 10000 K on
days 49 and 57, 7200 K on day 79 and 7200 K on day 89.  We
find that the black-body fits work best for $E_{B-V} = 0.15$ mag, consistent with
our estimate from \nai\ absorption ($0.17 \pm 0.06$ mag; SCL98).

\subsection{Elements identified}

\subsubsection{Hydrogen}\label{sec-h}

We detect the Balmer series from \ha\ to at least H11, and possibly
also members of the Paschen series from Pa$\delta$ to Pa11 (Fig.\
\ref{f-idents1}).  The profiles seem to be a combination of broad
emission line with $\mbox{BVZI} \gtsim 4000$ km s$^{-1}$ and a narrow
P-Cygni absorption component ($\sim 1000$ km s$^{-1}$).  The ratio of
narrow-to-broad emission component decreases from H$\alpha$ toward the
higher Balmer lines.  This indicates a nearly normal Balmer decrement
for the narrow component and a flat or possibly inverse Balmer
decrement for the broad component (Fig.\ \ref{f-profiletypes}).

A flat or inverted decrement suggests strong deviation from the
recombination case, related to collisions and radiative transitions in the
field of the photospheric radiation.  The observed line ratios suggest an
excitation temperature for hydrogen's excited levels of around
15000 K.  Two different thermalization mechanisms for hydrogen levels may
operate in this situation: the level populations may be controlled by
radiative transitions in the field of the external black-body photospheric
radiation, or thermalization may be primarily collisional. Both mechanisms
have been exploited to account for flat or inverted Balmer decrements
observed in cataclysmic variables \citep{elitzur83,williams88}. Both
thermalization mechanisms are plausible in SN~1994W as well.

\subsubsection{Helium}\label{sec-he}

We observe broad \hei\ lines with triangular emission profiles and
narrow P-Cygni features.  The \hei\ lines disappear after day 49
(Figs.\ \ref{f-spectra-dyn} and \ref{f-luminosities}; Table
\ref{tab-line-ids1}). \hei\ $\lambda$5876 and $\lambda$7065 appear only in
emission.  \hei\ $\lambda$4471 and $\lambda$6678 show weak absorption
features.  A high ratio $I(7065)/I(5876) \approx 0.5$ indicates a large
contribution of collisional and/or radiative excitation compared to
the purely recombination case.  Unfortunately, this fact cannot be
used as a straightforward indicator of electron concentration in the
line-emitting region.  Computations for an extended parameter set
\citep{an89} show that this particular ratio may be reached in a wide
range of concentrations ($n_e \approx 10^4-10^{14}$ cm$^{-3}$), with
pronounced dependence on the optical depth in \hei\ $\lambda$3889.

\subsubsection{Oxygen}\label{sec-o}

\Oi\ $\lambda7773$ is present with an absorption-dominated P-Cygni
profile from as early as day 18 to day 89. From day 79, we tentatively
identify \Oi\ $\lambda7002$ with a P-Cygni profile.  This line has not
to our knowledge been seen in a supernova spectrum before, but was
identified in the proto-planetary nebula Henize 401 \citep*{garcia99},
whose low-ionization spectrum bears some similarity to that of SN 1994W at
these phases.

\subsubsection{Other metals}\label{sec-metals}

\caii\ H \& K were strong in absorption at all epochs when we covered
them.  The near-infrared triplet lines were weak on day 49, but
strengthened dramatically by day 79 (Fig. \ref{f-idents1}).  This may
reflect an increase in the \caii/\caiii\ ratio with the drop in
radiation temperature at late epochs.

The \nai~D doublet's development is remarkable (Figs.\
\ref{f-spectra}, \ref{f-spectra-dyn} and \ref{f-profiletypes}), even
taking into account blending with interstellar absorption and in the
earlier spectra with \hei\ $\lambda5876$.  Up to day 49, a flat-topped
profile emerged as \hei\ faded.  From day 57 on, the strength of the
circumstellar P-Cygni absorption increased, presumably as a result of
increasing \nai\ ionization fraction.

Magnesium lines appear both in absorption and emission.  Narrow
absorption with weak P-Cygni emission is seen in \mgii\ $\lambda4481$
from day 21 to day 57.  We identify \mgii\ $\lambda$7877--7896 as the
source of the broad feature around 7890~\AA.  As an alternative
interpretation, we considered \fexi\ $\lambda$7892.  Strong \fexi\
might be expected to be accompanied by emission in \fex\
$\lambda6374$, and there may indeed be a broad feature underlying the
\Siii\ absorption lines to the blue of \ha.  However, this feature
does not persist to later epochs as the 7890~\AA\ feature does, nor
are any other high-ionization lines identified at any epoch.  In
addition, emission lines with similar widths to the 7890~\AA\ feature
are seen at wavelengths corresponding to \mgii\ transitions around
9230~\AA, and possibly at 8200~\AA\ and 10900~\AA.  Nevertheless, we
find it somewhat surprising that the broad emission in \mgii\
$\lambda\lambda$7877--7896 is not accompanied by similar emission
in \Oi\ $\lambda$7773.

The doublet of \Siii\ $\lambda\lambda$6347, 6371 is present from days
21 to 89, showing absorption-dominated P-Cygni profiles.  The lines
weakened relative to other features on days 79 and 89.  \Siii\
$\lambda6240$ therefore seems to be an unlikely contributor to the feature
at 6260~\AA\ (see below), which increases on these dates.

\scii\ emission lines at around 5530~\AA\ and 6250~\AA\ were
identified by \citet{pastorello02} in the spectrum of the Type IIn SN
1995G.  We identify both \scii\ $\lambda5527$ and $\lambda6246$ in our
spectra, and \scii\ may contribute to the feature at 6260~\AA.  In
particular, $\lambda5527$ shows some evidence of a broad emission
component (Figs.\ \ref{f-spectra-dyn} and \ref{f-idents1}).  Its
narrow P-Cygni core apparently persisted until day 121.

We identify a host of \feii\ lines (Fig.\ \ref{f-idents1}; Table
\ref{tab-line-ids1}), all of which show narrow P-Cygni profiles.  They may also
be accompanied by broad emission as in the case of other species, since the
spectra show hints of underlying broad emission in regions of the spectrum
where there are many \feii\ lines (Fig.\ \ref{f-spectra}).  By days 79 and
89, a few lines of forbidden and semi-forbidden \feii\ are identified.
These have emission-dominated P-Cygni profiles (Fig.\ \ref{f-idents1}).


\section{General picture} \label{sec-picture}

The previous discussion of spectroscopic and photometric data on SN~1994W
(SCL98) led to the conclusion that the narrow lines originate in a dense CS
envelope. However, at present, the spectroscopic and photometric results
cannot readily be incorporated into any existing model of a SN~II interacting
with a CS environment.  We therefore present first our qualitative view of
what we observe in the case of SN~1994W.  We emphasize some basic elements
of the physical picture that, in our opinion, are crucial for
understanding the phenomenon.

\subsection{The opaque cool dense shell}\label{sec-cds}

The smooth continuum of SN~1994W and its lack of broad absorption
lines is typical of most SNe~IIn at an early phase.  It is also
reminiscent of the early spectrum of SN~1998S \citep{leonard00}.  For
SN~1998S, the smooth continuum and the absence of broad absorption lines
was explained as an effect of an opaque cool dense shell (CDS) which
formed at the interface of the SN with its circumstellar medium
\citep{chugai01}.  In this situation, an opaque CDS is physically
equivalent to an expanding photosphere with a sharp boundary and
without an external, extended SN atmosphere.  In comparison, the CDS
formed in supernova remnants during the radiative stage
\citep{pikelner54} and in SNe interacting with a moderately dense CS
wind \citep{cf85} are optically thin in the continuum.

The large optical depth of the CDS requires a relatively large swept-up
mass.  This may be a natural consequence of shock break-out in SNe~II with
extended stellar envelopes \citep{grasberg71,fa77,bb93} and/or with
extended CS envelopes of unusually high density.  In the model of a
pre-supernova with initial radius $\sim10^{15}$ cm \citep{fa77}, a CDS with
a mass of $\sim 2~M_{\odot}$ forms and remains opaque for about 70 days.
We propose that a similar opaque CDS formed in SN~1994W.  This suggestion
finds support in the exceptionally bright peak of the light curve, $M_V \approx
-19.5$ mag (SCL98, their Fig.\ 5).  According to the theory of SN~II light
curves, a broad luminous maximum with $M \approx -(19-20)$ mag requires the
explosion of a very extended pre-supernova with an envelope radius of
$R_0 \approx 10^{15}$ cm \citep{grasberg71,fa77}.

\subsection{The broad-line region}\label{sec-broad}

The deceleration of the CDS in the extended pre-supernova envelope and in
the dense CS environment is accompanied by the Rayleigh-Taylor (RT)
instability \citep{fa77,chevalier82}.  The growth and subsequent
fragmentation of RT spikes in the dense CDS material results in the
formation of a narrow layer ($\Delta R/R \approx 0.1-0.15$) composed of dense
filaments, sheets and knots embedded in the rarefied hot gas behind the
forward shock \citep{cb95,be01}.  This mixed layer on top of the CDS is
likely responsible for the broad emission lines observed in SN~1994W.  If
this is the case, then the expansion velocity of the CDS ($v_{\rm s}$)
should be the same as the velocity of the broad line-emitting gas, i.e.,
$v_{\rm s} \approx 4000$ km s$^{-1}$.

A drawback of this picture of the broad-line formation is that the velocity
distribution of the line-emitting gas peaks at the CDS expansion
velocity $v_{\rm s}$, which means the expected line profile should be
boxy.  With the exception of the \nai\ profile on day 57, the broad
lines are instead triangular.

A triangular profile can be produced, however, if the spherically symmetric
velocity distribution of the line-emitting matter is close to $dM/dv \propto
v$.  Actually, given this mass-velocity spectrum and constant emissivity per
unit mass, and assuming that the velocity field is
spherically symmetric,  the luminosity distribution over the
radial velocity $u$ (in units of $v_{\rm s}$) is
\begin{equation}
\frac{dL}{du}\propto\int_{|u|}^{1}\frac{1}{v}\frac{dM}{dv}dv\propto(1-|u|)\,,
\label{eq-triang}
\end{equation}
which is indeed triangular.

The broad velocity spectrum of the dense gas in the mixed layer may
naturally arise in two plausible scenarios.  In the first scenario the CS
gas is clumpy. In this case, an ensemble of dense CS clouds engulfed by the
forward post-shock gas could provide a broad velocity spectrum of radiative
cloud shocks and cloud fragments \citep{chugai97} in the range of
1000--4000 km s$^{-1}$, where the lower limit is the CS velocity. This
model may be characterized as low-velocity clouds in a high-velocity flow.

Alternatively, dense high-velocity clumps interacting with the
low-velocity CS gas could produce a similar spectrum of fragments.  This
could occur if the forward shock is radiative, forming a thin
post-shock layer with $\Delta R/R < 0.1$.  In this situation, dense RT
spikes, formed at the previous stage of RT instability of the CDS, can
penetrate into the pre-shock zone.  The interaction of these
protrusions with the CS gas in the pre-shock zone leads to further
fragmentation and deceleration of fragments down to the velocity of
the CS gas.  As a result, a broad velocity spectrum of dense cool
fragments emerges in the velocity range of 1000--4000 km s$^{-1}$.

For this second possibility to hold, a high CS density $n$ is needed to
maintain the narrow width of the post-shock cooling region,

\begin{equation}
l_{\rm c}=\frac{m_{\rm p}v_{\rm s}^3}{32n\Lambda},
\label{eq-cool}
\end{equation}

\noindent where $m_{\rm p}$ is the proton mass, and $\Lambda$ is the
cooling function in the post-shock zone. To reach the required ratio
$l_{\rm c}/R \approx 0.1$ at radius $R = 10^{15}$ cm, one needs a CS
density of $n \approx 1.7 \times 10^9$ cm$^{-3}$ assuming
$\Lambda = 2 \times 10^{-23}$ erg s$^{-1}$ cm$^6$ and $v_{\rm s} = 4000$ km
s$^{-1}$.

Both scenarios for the velocity spectrum leave open the question of
why the velocity distribution of the line-emitting gas in the mixed
layer is close to $dM/dv \propto v$.  The following naive model seems
to provide a hint.

Let us consider the interaction of a high-velocity fragment with the CS
gas. In the rest frame of the initial fragment, the fragmentation process in
the rarefied CS flow with the velocity $\approx v_{\rm s}$ may be thought
of as a stripping flow with accelerating velocity $v_1$ and in which the
``radius'' $a$ of fragments progressively decreases as the velocity $v_1$
increases \citep*{klein94}.  Assuming mass conservation of fragments in
velocity space, $Q = dM/dt = (dM/dv)(dv/dt) = {\rm constant}$, the mass-velocity
spectrum of fragments may be expressed as

\begin{equation}
\frac{dM}{dv_1}=Q\left(\frac{dv_1}{dt}\right)^{-1}\,,
\label{eq-dvdt}
\end{equation}

\noindent
which is essentially determined by the acceleration $dv_1/dt$. The latter
can be approximately described as acceleration due to the drag force

\begin{equation}
 m\frac{dv_1}{dt}=\pi a^2\rho (1-v_1)^2\,,
\label{eq-drag}
\end{equation}

\noindent
where the velocity is in units of $v_{\rm s}$, $m=(4\pi/3)a^3\rho_{\rm
f}$ is the mass of fragment of the density $\rho_{\rm f}$, and
$\rho$ is the density in the rarefied ambient flow.  To determine the
acceleration we must specify the relation between fragment size and
velocity $v = 1-v_1$ relative to the CS gas.  We assume  that the
fragment size for each new generation of cloudlets is tuned to the
Kolmogorov turbulent cascade with spectrum $v \propto
a^{1/3}$.  With this relation,  equation (\ref{eq-drag})
leads to $dv_1/dt \propto 1/v$.  The latter combined with
equation (\ref{eq-dvdt}) leads to a mass-velocity spectrum
$dM/dv \propto v$.

A similar analysis carried out for the case of a CS cloud in the SN flow
produces a quite different mass-velocity spectrum, $dM/dv \propto (1-v)$.
This spectrum also results in a broad line but with zero slope at $|u|=1$
and a narrow logarithmic peak: $dL/du \propto (|u|-1-\ln\,|u|)$. While this
profile resembles the observed H$\alpha$ profile, it is at odds with the
triangular profile of \hei\ lines.  Below (Section \ref{sec-haev}) we argue
against a large contribution of CS clouds in the formation of H$\alpha$.

 \begin{figure}
  \centering
  \vspace{8cm}
  \includegraphics{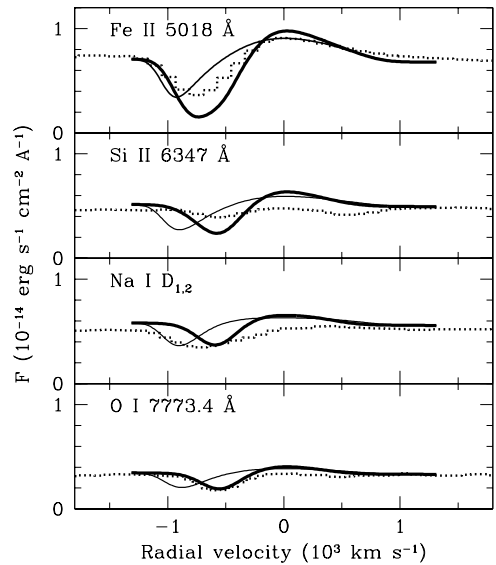}
  \caption []{Circumstellar lines  on day 79. Overplotted on the
  observational data ({\em dotted line}) are the models for
  homologous expansion ({\em thick line}) and the constant-velocity
  case ({\em thin line}).
    }
  \label{f-lte}
  \end{figure}

\subsection{The circumstellar envelope}\label{sec-csenv}

\subsubsection{Expansion kinematics}\label{sec-kin}

The radial velocities of narrow absorption features in SN 1994W are
strikingly persistent between days 18 and 89.  This implies that the
expansion velocity of the CS envelope ($\sim 1000$ km s$^{-1}$) has a
pre-explosion origin.  An alternative possibility --- acceleration of
a slow wind by SN radiation (see also SCL98) --- seems unlikely, since
it would require unrealistic fine tuning between the time-dependent
luminosity and absorption (scattering) coefficient to arrange the
persistence of the velocity of accelerated gas.

Two extreme options are conceivable for the CS envelope's
pre-explosion kinematics: constant-velocity flow and homologous
expansion ($u \propto r$).  The former might be the result of either
continuous wind outflow or a succession of ejection events with
similar ejecta velocities.  Homologous expansion, on the other hand,
might be produced by a single explosive ejection.  The latter
mechanism, though exotic, was in fact proposed by \citet{grasberg86}
to account for the narrow CS lines in SN~1983K.

\subsubsection{Envelope size}\label{sec-size}

The light curve of SN~1994W is characterized by a broad maximum around
day 30 and a subsequent plateau, terminated by a sudden drop
in luminosity at $t_{\rm d} \approx 110$ d (SCL98, their Fig. 5).
Assuming a CDS expansion velocity of $v_{\rm s} = 4000$ km s$^{-1}$,
we find that during this period the CDS sweeps up the CS gas within
the radius $R_{\rm d}=v_{\rm s}t_{\rm d} \approx 4 \times 10^{15}$
cm.  Although this estimate refers to  $t=110$ d, we adopt
it as a rough estimate for earlier epochs as well.

We now argue that this radius should coincide with the outer radius of
the dense CS envelope, $R_{\rm cs} \approx R_{\rm d}$.  If instead
$R_{\rm cs} > R_{\rm d}$, then the light curve of SN~1994W should not
have shown such a steep drop, since the CS interaction would have
augmented the luminosity at the luminosity decay phase.  Equally, if
$R_{\rm cs} < R_{\rm d}$, the narrow lines would have disappeared at
epoch $t < t_{\rm d}$, which was not observed.  These arguments,
therefore, provide strong evidence that the extent of the dense CS
envelope is $R_{\rm cs} \approx R_{\rm d} \approx 4 \times 10^{15}$ cm.
Beyond this radius, the CS density presumably drops steeply, as indicated
both by the broad-band light drop after day 100 and the fast decline
of the H$\alpha$ luminosity (Fig.\ \ref{f-luminosities}).  At larger
radii, X-ray observations suggest that other density enhancements may
be present (Section \ref{sec-discussion}; \citealt{schlegel99}).

\subsubsection{Electron-scattering wings and circumstellar density}\label{sec-wings}

We attribute the extended smooth wings observed in H$\alpha$ and
H$\beta$ between days 18 and 89 to the effect of Thomson scattering on
thermal electrons participating in the bulk expansion of the CS
envelope
\citep{chugai01}.  Based on previous computations of Thomson scattering
effects in SN~1998S, and the strength of the H$\alpha$ wings in SN~1994W on
day 57 (when blending with \hei\ $\lambda$6678 was negligible), we believe
that the optical depth of the CS envelope to Thomson scattering ($\tau_{\rm
T}$) at this stage must have been close to unity.

Adopting a photospheric radius $R_{\rm p} \approx v_{\rm s}t = 2\times
10^{15}$ cm on day 57 (where $v_{\rm s}\approx 4000$ km s$^{-1}$) and an
outer radius of the CS envelope $R_{\rm cs}=4\times10^{15}$ cm, we
obtain an estimate of the average electron concentration in the CS
envelope assuming $\tau_{\rm T}\approx 1$:

\begin{equation}
 n_e\approx \frac{\tau_{\rm T}}
 {\sigma_{\rm T}(R_{\rm cs}-R_{\rm p})}\approx 10^9\;\mbox{cm}^{-3}.
\label{eq-ne}
\end{equation}

\noindent
This value, taken together with the forward shock-wave velocity ($v_{\rm
sh} \approx 3000$ km s$^{-1}$), after correction for the CS velocity, provides
an estimate of the density of the CDS material in the broad-line region
($n_{\rm cds}$).  The pressure equilibrium condition, with sound speed in
the CDS matter of $c_{\rm s} \approx 15$ km s$^{-1}$, results in $n_{\rm
cds} \approx n(v_{\rm sh}/c_{\rm s})^2 \approx 4\times10^{13}$ cm$^{-3}$.  This
tremendous density provides a natural reason for the strong collisional
thermalization of the broad component indicated by the observed inverse
Balmer decrement (Section \ref{sec-h}).

\subsubsection{Circumstellar density from narrow lines}\label{sec-narrow}

The strength of narrow subordinate lines indicates a high density in
the CS envelope, as noted by SCL98.  The simplest density estimate may
be taken from the condition that the optical depth in a line of \feii,
for example $\lambda$5018, is on the order of unity.  This follows
from the relative intensity of the absorption component of this line
($\sim 0.5)$.  Assuming that level populations of \feii\ obey the
Boltzmann distribution for $T=10^4$ K, one obtains for solar Fe/H a
hydrogen concentration of $n \approx 3 \times
10^7v_8r^{-1}_{15}f_2^{-1}$ cm$^{-3}$.  Here $r_{15}$ is the linear
scale of the line-forming zone in units of $10^{15}$ cm, $v_8$ is the
velocity dispersion in units of $10^8$ cm s$^{-1}$, and $f_2$ is the
ionization fraction of \feii.  Given the scale of the CS envelope
($\sim 4 \times 10^{15}$ cm), we thus have a lower limit $n > 10^7$
cm$^{-3}$.  This lower limit should be considered as a revised version
of the estimate reported in SCL98.

A somewhat more elaborate estimate of the density may be obtained
using several different lines and assuming a quasi-local thermodynamic
equilibrium (quasi-LTE) approximation,
i.e., the Saha-Boltzmann equations corrected for both geometrical dilution
$W$ and dilution of the photospheric brightness $\xi$.  The latter
defines the photospheric brightness through the black-body brightness
$I_{\nu}=\xi B_{\nu}(T)$.  We adopt a simple density distribution in
the CS envelope: a plateau with a steep outer drop

\begin{equation}
n=n_0/[1+(r/R_{\rm k})^{12}]\;,
\label{eq-den}
\end{equation}

\noindent where the cutoff radius $R_{\rm k} \approx R_{\rm
cs} = 4 \times 10^{15}$ cm.  We consider two cases for the CS kinematics:
free expansion ($v=hr$), with outer velocity at $r=6\times 10^{15}$ cm
equal to $v_{\rm b}=1100$ km s$^{-1}$, and constant-velocity flow with
$v=1100$ km s$^{-1}$.  The photospheric radius on day 79 is estimated
as $R_{\rm p}\approx v_{\rm s}t\approx 2.7\times10^{15}$ cm, where
$v_{\rm s}=4000$ km s$^{-1}$, while the temperature is taken to be
$T_{\rm p}=7200$ K according to the value found from the black-body
fit to the continuum (Section \ref{sec-continuum}). The parameter
$\xi$ in this case is found to be 0.26.

In Fig.\ \ref{f-lte} we show calculated profiles of \feii\
$\lambda$5018, \Siii\ $\lambda$6148, the \nai\ D$_{1,2}$ doublet and
the \Oi\ 7773 \AA\ triplet, for solar abundance and hydrogen
concentration $n_0=1.5\times10^{9}$ cm$^{-3}$.  In the case of \nai\
we have for the sake of simplicity ignored the doublet structure,
which explains why our model line is narrower than the observed one.
This density seems to be the optimal one: a factor of 1.5
lower density results in too weak \Oi\ and \nai\ lines, while a factor
1.5 higher density makes all the lines too strong.

Although neither kinematic model produces an excellent fit, it is
clear that free expansion kinematics better predicts the positions of
the absorption minima and, therefore, is preferred compared to the
constant-velocity case.  The differences between the modelled and
observed line profiles in the free-expansion case may be related to
the omission of Thomson scattering and the simplicity of our
quasi-LTE model.

 \begin{figure}
  \centering
  \vspace{8cm}
  \includegraphics{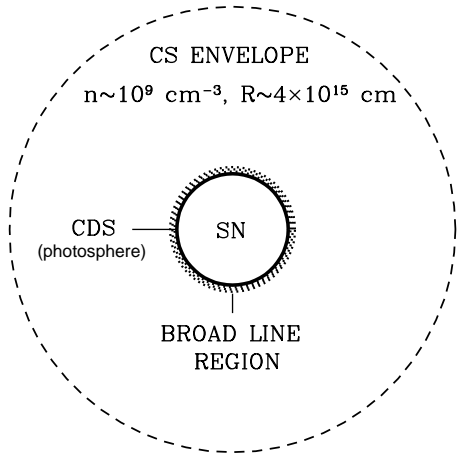}
  \caption []{A sketch of SN~1994W at the epoch around day 30.
We show the principal structural elements involved in the formation
of the spectrum. The SN ejecta are bounded by an opaque cool dense shell
(CDS), which is responsible for the continuum radiation. The broad-line
region is a narrow mixing layer attached to the CDS, composed of
Rayleigh-Taylor fragments of the CDS matter and possibly of shocked CS
clouds.  The SN ejecta expand into a dense CS envelope with Thomson
optical depth on the order of unity.  The CS envelope is responsible
for both narrow lines and the extended Thomson wings seen in
H$\alpha$.
  }
  \label{f-cart}
  \end{figure}

\subsection{Overview of the qualitative model}\label{sec-overview}

We summarize the main results of the qualitative analysis in a cartoon
(Fig.\ \ref{f-cart}). It shows the basic structural elements of SN~1994W
which we believe are responsible for the formation of the optical spectrum
when the supernova's luminosity is high.  The ejecta expand with a velocity
of $\sim 4000$ km s$^{-1}$ into an extended ($\sim 4\times10^{14}$ cm) CS
envelope.  The characteristic density of the CS envelope is $\sim 10^9$
cm$^{-3}$.  The SN ejecta are enshrouded by the opaque CDS within which the
photosphere resides during most of the luminous phase.  The opaque CDS
precludes the formation of absorption lines from the ejecta.  The CDS
presumably forms primarily during shock break-out and is subsequently
maintained by both CS material swept up by the radiative forward shock
and SN material swept up by the reverse shock.  These shocks are not shown
in Fig. \ref{f-cart} but they are presumably located at the distances $\Delta R<
0.1R$ away from the outer and inner edges of the CDS.

Attached to the CDS is a layer populated by dense ($n \approx
4\times10^{13}$ cm$^{-3}$) fragments supplied by the RT instability of
the dense CDS and, possibly, by radiative shocks in CS clouds.  This
inhomogeneous layer of dense material is the primary site of the
broad emission lines with the characteristic velocity of the line-emitting
gas, 4000 km s$^{-1}$ (e.g., in \hei).

The CS envelope, which in turn expands with a velocity of $\sim 1000$ km
s$^{-1}$, is responsible for the narrow lines.  Thomson scattering in the CS
envelope results in the emergence of broad emission-line wings, which are
most apparent in the strong lines such as H$\alpha$ and H$\beta$.  Because
of the large expansion velocity, the red wing is stronger than the blue wing.

\section{A model for H$\alpha$}\label{sec-line}

In order to confirm and refine the above qualitative picture of the
formation of the SN~1994W spectrum, we have modelled the H$\alpha$ line
profile.  In many respects our approach repeats the analysis of SN~1998S by
\citet{chugai01}, which involved calculations of the CDS dynamics and Monte
Carlo computations of the emergent lines formed outside the CDS.  Here we
use an updated version of this model, which now includes calculations of CS
hydrogen ionization, electron temperature and H$\alpha$ emissivity.  In the
previous model both of the latter were set rather arbitrarily.

\subsection{The model}

\begin{figure}
  \centering
  \vspace{9cm}
  \includegraphics{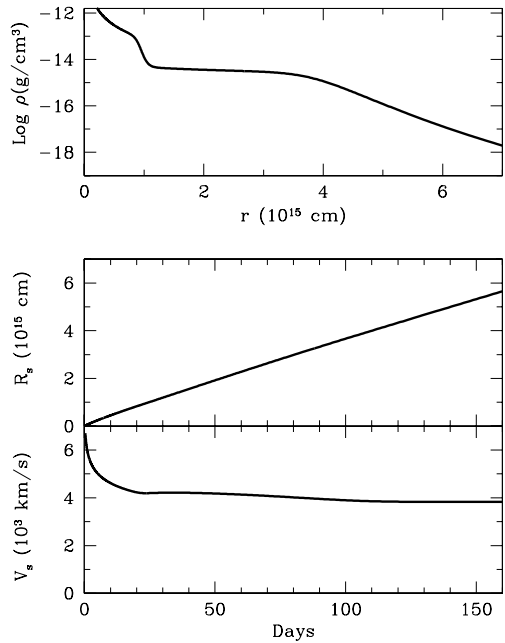}
  \caption[]{The CS density ({\em top panel}) and
  evolution of the radius and velocity of the cool
  dense shell ({\em lower panels}).
  }
  \label{f-dyn}
  \end{figure}

The CDS dynamics are calculated numerically in the thin-shell
approximation \citep{chevalier82}.  We assume that the SN initially
expands homologously ($v\propto r$) with density distribution
$\rho=\rho_0/[1+(v/v_{\rm k})^9]$, where the parameters of the
distribution are defined by the ejecta mass $M$ and kinetic energy
$E$.  At radius $R_0=1000~R_{\odot}$ (a rather arbitrary value) the SN
ejecta begin to interact with the extended stellar envelope.  The
density of the extended stellar envelope is adjusted in our model by
the requirement that a strong initial deceleration of the swept-up
thin shell must result in an expansion velocity of $\sim 4000$ km
s$^{-1}$ on day 30 when the photospheric radius reaches
$\sim10^{15}$ cm. The CS envelope is attached to the stellar envelope
at $r \approx 10^{15}$ cm.  A constant pre-shock velocity (1000 km s$^{-1}$)
is assumed for the CS gas.  This is a reasonable approximation for the
homologous expansion, bearing in mind additional radiative
acceleration.  For the SN mass and energy we adopt $M=8~M_{\odot}$ and
$E=10^{51}$ erg --- both values are close to the parameters of
the light-curve models B and C of \citet{fa77}.  We find that the dynamical
model is nearly the same for SN mass in the range $7-9~M_{\odot}$.
The evolution of the CDS radius and velocity is shown in Fig.\
\ref{f-dyn} for the density profile which is also consistent with the
H$\alpha$ model.  In fact, the optimal density is a compromise between
the requirements imposed by the H$\alpha$ profile evolution, the CDS
velocity, and the bolometric luminosity at the final stage of the
light-curve plateau.  The density of the CS envelope is nearly flat
($\rho \propto r^{-0.4}$) in the range $10^{15}< r <4 \times 10^{15}$ cm.

The ionization of the CS envelope is produced by both X-ray emission
from the radiative forward shock wave and by photospheric
radiation.  The latter primarily operates via hydrogen photoionization
from the second level.  The X-ray luminosity of the forward shock wave
is calculated as

\begin{equation}
L_{\rm x}=2\pi r^2\rho (v_{\rm s}-u_{\rm ps})^3,
\label{eq-lsh}
\end{equation}

\noindent where $\rho$ is the pre-shock density, while
the pre-shock velocity $u_{\rm ps}$ is the superposition of the
pre-explosion velocity and the accelerated velocity term.  Half of
this luminosity is directed toward the CDS where it is reprocessed
into continuum radiation.  This component is not treated explicitly in
the model.  The other half of the X-ray luminosity of the forward
shock is emitted outward and is partially absorbed by the CS gas.  The
X-ray spectrum of the forward shock is approximated as
$F_E=CE^{-q}\exp\,(-E/T_{\rm s})$, where $T_{\rm s}$ is the
temperature of the forward shock defined by the shock velocity $v_{\rm
s}-u_{\rm ps}$.  The spectral power index $q=0.5$ roughly describes
both the Gaunt factor and the contribution of X-ray lines in the
low-energy band
\citep{terlevich92}.  The soft X-rays from the reverse shock at the
epoch under consideration are fully absorbed by the CDS (cf. Fig. 9
of \citealt*{flc96}) and by the SN ejecta, and are reprocessed into the optical.

The energy of the X-rays absorbed in the CS envelope is shared between
heating, ionization and excitation.  Excitation and ionization by the
photospheric radiation are also taken into account.  The electron
temperature $T_{\rm e}$ in the CS envelope is determined from the
energy balance between heating from ionization by X-rays and
photospheric radiation and cooling due to hydrogen. This approximation
is quite reasonable in the temperature range of $(1-2)\times 10^4$ K.
The hydrogen atom is treated in the two-level-plus-continuum
approximation.  The degree of ionization we obtain for hydrogen in the
CS envelope lies typically in the range $0.8-0.95$.  The H$\alpha$
emissivity is due to recombination (Case B) and collisional
excitation. Collisional de-excitation of H$\alpha$ is included in the
computation of the net emissivity.  To allow for uncertainties in the
flux, model, distance and reddening we use a fitting parameter
$C_{\rm n}$ for narrow H$\alpha$.  This parameter is a multiplicative
factor in the H$\alpha$ emissivity in the CS envelope.

The broad-line region (Fig. \ref{f-cart}) is described by an (on
average) homogeneous layer, $R_{\rm s}<r<R_1$, where $R_1/R_{\rm
s}=1.15$ and the distribution of the line luminosity $dL/dv\propto v$
for $v_1<v<v_{\rm s}$, where $v_1$ is the minimal velocity adopted to
be equal to the pre-shock CS gas velocity.  We express the average
emissivity of the broad component in terms of the pre-shock
emissivity, multiplied by a fitting factor $C_{\rm b}$, which is
restricted by the requirement that the H$\alpha$ intensity from the
broad-line region cannot exceed the black-body value for the local
electron temperature of the CS gas in the pre-shock zone.

We assume free expansion pre-explosion kinematics ($u=hr$), though we
have also explored the constant-velocity case.  The acceleration by
the SN radiation is described as an additional CS velocity term
$u_{\rm a}=u_{\rm 0}(R_{\rm s}/r)^2$, where $u_{\rm 0}$ is a free
parameter.

\subsection{Thomson scattering and the broad component}\label{sec-broadth}

\begin{figure}
  \centering
  \vspace{6.8cm}
  \includegraphics{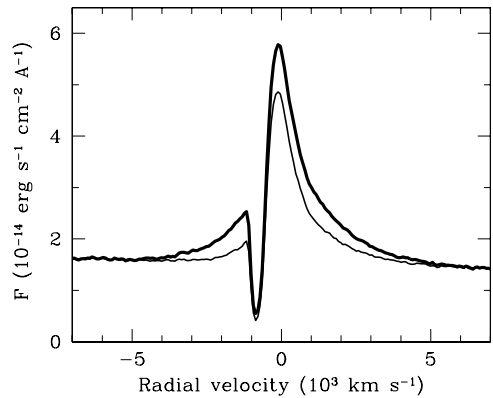}
  \caption[]{A typical model of H$\alpha$ on day 31 ({\em thick line}),
  showing the contribution of the narrow component ({\em thin line}).
       }
  \label{f-broad}
  \end{figure}

\begin{figure}
  \centering
  \vspace{7.5cm}
  \includegraphics{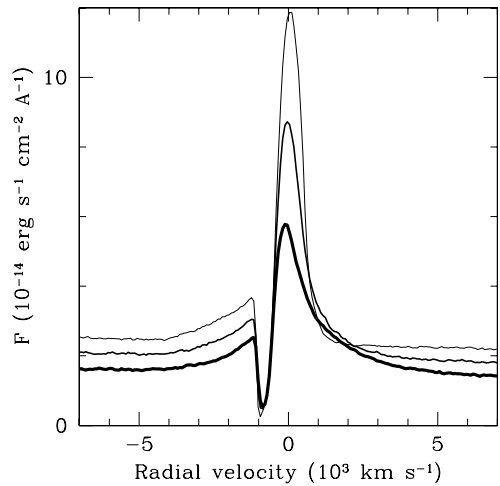}
  \caption[]{The same model as in Fig. \ref{f-broad} but for different
  Thomson optical depths.  Lines with increasing thickness show cases
   $\tau_{\rm T}=0$, 1 and 2.8, respectively.
     }
  \label{f-thomson}
  \end{figure}

In the H$\alpha$ line profile, the broad emission component and
electron-scattering wings overlap and at first sight cannot be
disentangled unambiguously.  Nevertheless, we found that there is not
much freedom in the decomposition procedure.  In Fig. \ref{f-broad} we
show a model narrow-line component on day 31, with the broad component
turned off.  To emphasize the different behaviours of the broad
component and the electron-scattering wings, we show in Fig.\
\ref{f-thomson} models with different Thomson optical depth (0, 1 and
2.8) but with otherwise similar parameters. For optical depth zero,
the modelled broad component is strongly skewed toward the blue because
of occultation by the photosphere. This asymmetry decreases as
$\tau_{\rm T}$ increases and Thomson scattering of the line emission
in the expanding CS envelope produces an increasingly strong red wing.
These different behaviours of the broad line and electron-scattering
wings permit us to disentangle the contribution of the two line
components in H$\alpha$.  In passing, we note that the continuum level
also depends on $\tau_{\rm T}$, since the Thomson optical depth
affects the escape probability of photons emitted by the photosphere.

\subsection{H$\alpha$ and the expansion law}\label{sec-explaw}

\begin{figure}
  \centering
  \vspace{9cm}
  \includegraphics{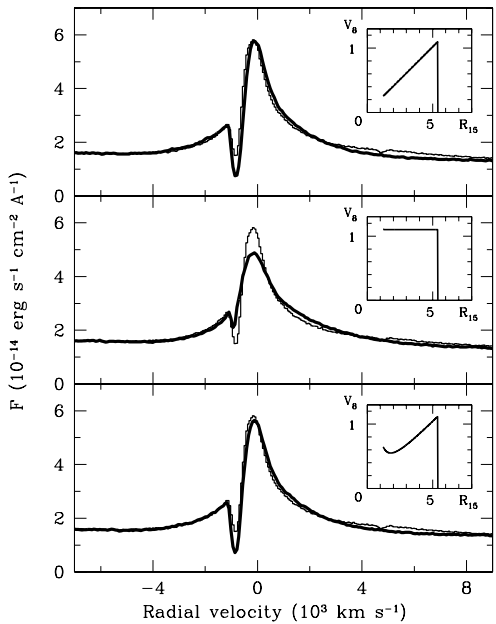}
  \caption[]{H$\alpha$ profile on day 31 for different
  velocity distributions (shown in the inset in each panel).
  Overplotted on each observed profile ({\em thin line})
   are models for CS envelope kinematics displayed in insets.

  }
  \label{f-kinem}
  \end{figure}

The sensitivity of the model line profile to the kinematics of the CS
envelope is demonstrated in Fig.\ \ref{f-kinem}.  The plot shows
H$\alpha$ profiles on day 31 computed for similar model parameters ---
all that differs is the velocity distribution.  For $\tau_{\rm
T}=2.8$, homologous expansion without post-explosion radiative
acceleration fits the observations quite well.  The constant-velocity
case is much less successful: it produces an emission component which
is too broad, and absorption which is too shallow
(Fig. \ref{f-kinem}).  This result, taken together with the results of
the narrow CS line modelling (Fig. \ref{f-lte}), argues against the
constant-velocity case.  We checked a model with constant velocity and
pre-shock acceleration and found, unsurprisingly, that this gives even
worse agreement.  The homologous model, on the other hand, works just
as well when combined with post-explosion acceleration.  In the bottom
panel of Fig. \ref{f-kinem} we show such a model profile characterized
by an amplitude of accelerated velocity term of $u_{\rm a}=400$ km
s$^{-1}$.

Although homologous expansion kinematics is certainly preferred, this
model shows a rather deep absorption component.  This contradiction is
not related to the finite resolution since we have smoothed the model
profile by a Gaussian with appropriate width.  An explanation for the
large absorption strength in the model might be hidden in a possible
deviation of the kinematics from homologous expansion, some clumpiness
of the CS envelope, or lower hydrogen excitation in the outer layers
of the CS envelope than the model predicts.

 \begin{figure}
  \centering
  \vspace{8cm}
  \includegraphics{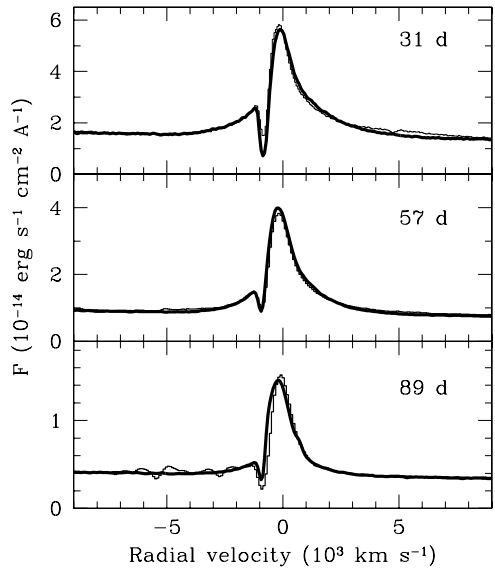}
  \caption[]{H$\alpha$ at different epochs.
   Model profiles ({\em thick line})  are
  overlaid on the observations ({\em thin line}).
   }
  \label{f-haev}
  \end{figure}

\subsection{Modelling the H$\alpha$ evolution}\label{sec-haev}

We have calculated H$\alpha$ for three epochs (31, 57 and 89 days past
explosion) using the density distribution of the CS envelope
($r>10^{15}$ cm), and the CDS radius and velocity as shown in Fig.\
\ref{f-dyn}.  The density parameter $w=4\pi r^2\rho$ at $r=10^{15}$ cm
on day 31 is $w=7.5\times10^{16}$ g cm$^{-1}$.  Free expansion
kinematics is assumed with a boundary velocity of $1100$ km s$^{-1}$
at $r=5.4\times10^{15}$ cm on day 31.  This implies an age for the CS
envelope of $t_{\rm cs}=1.5$ yr when the SN explodes.  For other
values of $t_{\rm cs}$ counted from the moment of envelope ejection,
we rescale according to $r\propto t_{\rm cs}$ and $\rho\propto
1/t_{\rm cs}^3$.

The computed profiles are shown in Fig. \ref{f-haev}.  In Table
\ref{t-par} we list the corresponding parameters: photospheric (CDS)
radius, photospheric temperature, brightness dilution, CDS velocity,
post-explosion velocity increase, and emissivity fitting factors for
narrow and broad components.  The last column displays the calculated
Thomson optical depth outside the CDS.  On days 31 and 57 we use the
approximation of an opaque photosphere.  On day 89 we found, however,
that this approximation predicts too weak a red part of the profile.
The model agrees better with the observations if we suggest that at
this late phase the CDS is semi-transparent.  To describe this effect
in a simple way we allow a photon striking the photosphere to
cross it and escape into the CS medium with a finite probability,
which we found should be close to $p=0.3$.  This assumption is
qualitatively consistent with the fact that the photospheric
brightness on day 89 is diluted ($\xi=0.2$) for the optimal continuum
temperature. The models for all three epochs demonstrate
satisfactory fits to the data.  On day 31, the model is consistent
with the presence of broad \hei\ $\lambda$6678 in the red wing of
H$\alpha$, while on day 57 the model fit in the red wing does not
leave any room for the \hei\ line.  This behaviour is consistent with
the weakening of other \hei\ lines at this time.

We calculated the H$\alpha$ component from the CS envelope in a
unified way with a minimum number of free parameters.  For this
reason, the fact that the tuning parameter of the narrow-line
intensity ($C_{\rm n}$) is close to unity (Table \ref{t-par}) means
that within the uncertainties ($\sim 20$ per cent) the electron distribution
in the CS envelope recovered from the Thomson scattering effects is
also consistent with the luminosity of the narrow H$\alpha$ component.
Because the line luminosity depends on the filling factor as $L\propto
1/f$ while the Thomson optical depth does not, we conclude that the
filling factor in the CS envelope is close to unity in at least most
of the line-forming zone.

This last observation is of crucial importance for distinguishing between
the two scenarios for the formation of the broad-line region (Section
\ref{sec-broad}), and indicates that CS clouds do not play a significant
r{\^o}le in this process.

We have studied cases with different density power-law indices $s$ (defined
via $\rho\propto r^{s}$) in the flat part of the CS envelope.  In the
case $s=0$, the tuning parameter ($C_{\rm n}$) must systematically
decrease by a factor of 1.25 to fit the profile.  For
$s=-1$, $C_{\rm n}$ should systematically increase with time by a
similar factor. Bearing in mind the uncertainties in the modelling
and fluxing, we do not rule out that the density gradient lies in the
range $-1\leq s\leq 0$.

\subsection{Implications for the origin of the circumstellar envelope}\label{sec-impl}

The model for the H$\alpha$ evolution presented above for the CS
envelope outside $r \approx 10^{15}$ cm argues for a total mass of
$M_{\rm cs}\approx 0.4~M_{\odot}$ and kinetic energy of $E_{\rm
cs} \approx 2 \times 10^{48}$ erg.  Combined with the derived kinematic
age of the CS envelope $t_{\rm cs} \approx 1.5$ yr, the average kinetic
luminosity of the mass-loss mechanism responsible for the ejection of
the CS envelope then becomes $L\approx E_{\rm cs}/t_{\rm cs} \approx
4\times10^{40}$ erg s$^{-1}$, and the average mass-loss rate is
$\sim 0.3~M_{\odot}$ yr$^{-1}$.

The kinetic luminosity we derive here is enormous compared to stellar
values.  It exceeds by two orders of magnitude the radiative
luminosity of a pre-supernova with a main-sequence mass of $\leq
20~M_{\odot}$.  This certainly rules out a superwind as the mass-loss
mechanism responsible for the CS envelope around SN~1994W.  The CS
envelope around SN~1994W must have been born as a result of a rather
violent mass ejection initiated by some energetic explosive event in
the stellar interior ($E \geq 2\times 10^{48}$ erg) at $\sim 1.5$ yr
prior to the SN outburst.


\begin{table*}
  \caption{Parameters of the H$\alpha$ evolution models}
  \bigskip
  \begin{tabular}{lcccccccc}
  \hline

Day   & $R_{\rm p}$  &  $T$ & $\xi$ &  $v_{\rm s}$  & $u_{\rm a}$  &  $C_{\rm b}$ &
      $C_{\rm n}$  & $\tau_{\rm T}$ \\

      & ($10^{15}$ cm) &  (K)  &  &  (km s$^{-1}$)  &  (km s$^{-1}$)  &             &
                   &             \\

\hline

31 &   1.2         &  16000 & 1 & 4220        &   400         &  35       &
         1.15     &  2.8 \\
57 &   2.2          &  7760 & 1 & 4160        &   400          &  1       &
         1    &  2.0 \\
89 &   3.3         &  7000 & 0.2 & 3920        &   170          &  0.5       &
         1.15     &  0.54 \\
\hline
\end{tabular}
\label{t-par}
\end{table*}

\section{Light-curve modelling} \label{sec-hydro}

To model the broad-band photometric light curves, we use the
multi-energy group radiation hydrodynamics code {\sc stella}
\citep{blin93j,blin87a}.  In the current work, {\sc stella}
solves time-dependent equations for the angular moments of intensity
averaged over fixed frequency bands, using 200 zones for the
Lagrangian coordinate and up to 100 frequency bins with variable
Eddington factors.  The transfer of gamma rays from radioactive decay
is calculated using a one-group approximation for the non-local
deposition of the energy of radioactive nuclei.  Except for the latest
phases, the gamma-ray deposition is not important for SN~1994W.  In
the equation of state, LTE ionizations and recombinations are taken
into account, but radiation is not assumed to be in equilibrium with
matter.  The effect of line opacity is treated as an expansion opacity
according to the prescription of \citet{eastpin}.  In
comparison with previous work with {\sc stella}, here we use an
extended spectral line list, the same as in the discussion of SN 1998S
by \citet{chugai02}.

\subsection{Pre-supernova models}

SCL98 used analytical fits \citep{ln83,ln85,popov93}
to estimate the pre-SN radius, mass and explosion energy in SN 1994W.
However, those fits
are only valid for typical SNe II-P, and from the discussion above we know that
SN~1994W was instead dominated by circumstellar interaction.   To determine
the pre-supernova parameters for such an unusual event, we therefore need to
do detailed numerical modelling.  This can be done only
for a subset of the pre-supernova parameters, so we used a
cut-and-trial approach, guessing the initial conditions and computing
light curves in order to meet the constraints posed by the photometry
and the spectral analysis of Section \ref{sec-line}.  In the absence
of an  evolutionary model, we built a sequence of several dozens of
non-evolutionary models, converging finally on a reasonably good fit to observations.
Our initial models are constructed in hydrostatic equilibrium for the
bulk mass, while the outer layers mimic the structure of the
circumstellar envelope.

For a given mass $M$ and radius $R_0$, we obtain a model in mechanical
equilibrium assuming a power-law dependence of temperature on density
\citep{nadraz86,bb93}:

\begin{equation}
T\propto\rho^\alpha \;.
\label{eqmm}
\end{equation}

\begin{figure}
  \centering
  \vspace{8cm}
  \includegraphics{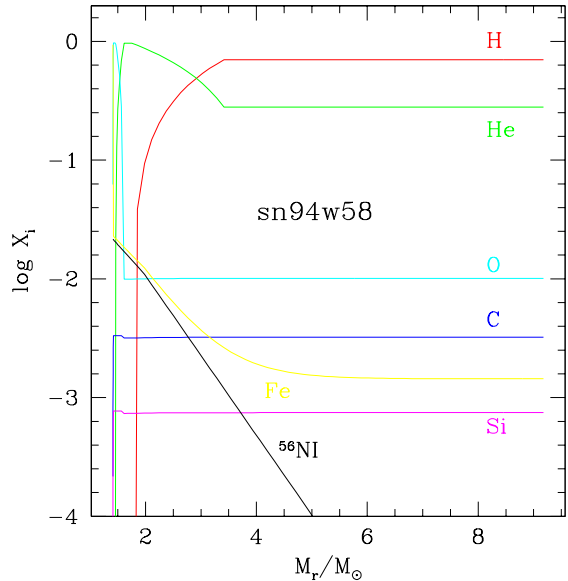}
  \caption[]{Composition as a function of interior mass,
$M_{\rm r}$, for the most abundant elements in
the pre-supernova model sn94w58.
The mass cut is at $M_{\rm c}=1.41~M_{\odot}$.
  }
  \label{chem}
  \end{figure}

\noindent
The hydrostatic configuration thus obtained would be close to a
polytrope of index $1/\alpha$ if it were chemically homogeneous and
fully ionized.  The difference from a polytropic model arises due to
recombination of ions in the outermost layers and non-uniform
composition (Fig.~\ref{chem}).

\begin{figure}
  \centering
  \vspace{8cm}
  \includegraphics{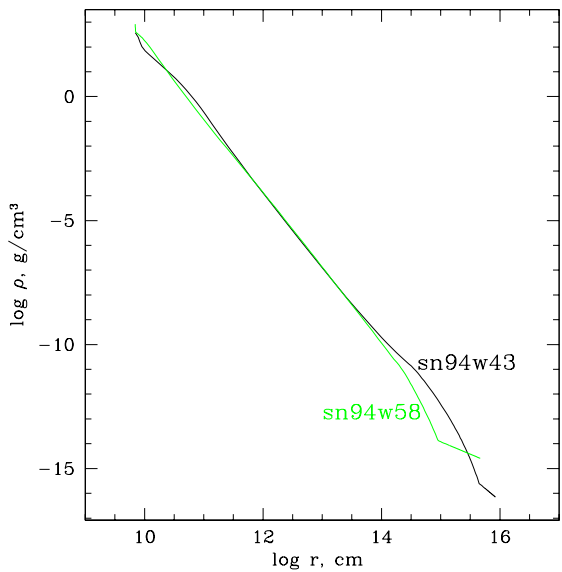}
  \caption[]{Density as a function
of the radius $r$ in the pre-supernova models sn94w43 and sn94w58.
  }
  \label{rhor}
  \end{figure}

\begin{figure}
  \centering
  \vspace{8cm}
  \includegraphics{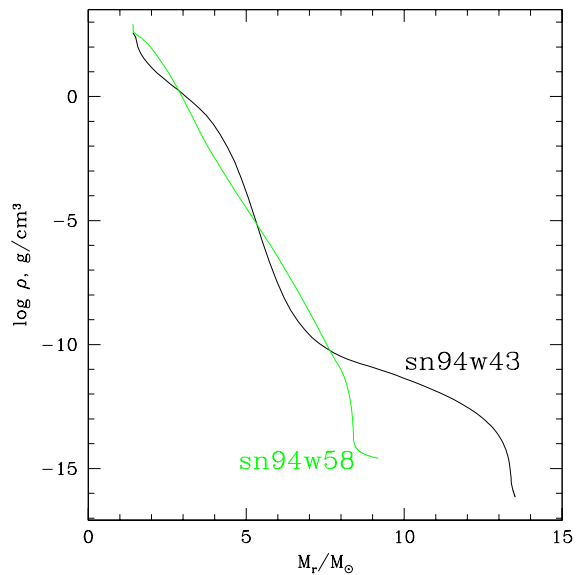}
  \caption[]{Density as a function of the interior mass ($M_{\rm r}$)
in the pre-supernova models sn94w43 and sn94w58.
The mass cut is at $M_{\rm c}=1.41~M_{\odot}$.
  }
  \label{rhom}
  \end{figure}

In the centre of this configuration, at the mass cut of the collapsing
core, we assume a ``point-like'' gravitating hard core (with numerical
radius $R_c = 0.1 R_\odot$ --- much larger than a real core, but much
smaller than the radii of mesh zones involved in our hydrodynamic
simulations).  The density structure found in this way is shown in
Figs.~\ref{rhor} and \ref{rhom}.

The parameters of the models are given in Table \ref{hydro}.  The
first column is the model label, which is followed by the mass of SN
ejecta, pre-supernova stellar radius, $^{56}$Ni mass, power-law index
$\alpha$, density of the CS envelope ($\rho_{15}$) at the radius
$r=10^{15}$ cm, power-law index ($p$) of the CS envelope density
distribution $\rho\propto r^{-p}$, the outer radius of the CS
envelope, and the kinetic energy at infinity in units of foe (1~foe =
$10^{51}$ erg).


\begin{table*}
\centering
 \begin{minipage}{140mm}
  \caption{Parameters of hydrodynamical models}
  \bigskip
  \begin{tabular}{lcccccccc}
  \hline
Model    & $M_{\rm ej}$\footnote{pre-supernova mass $=M_{\rm ej}+1.41~M_{\odot}$}  &   $R_0$         &  $M_{\rm Ni}$  &
$\alpha$  &   $\rho_{15}$    &   $p$   &  $R_{\rm w}$  &  $E_{\rm kin}$\footnote{kinetic energy at infinity} \\
         & ($M_{\odot}$) & ($10^4~R_{\odot}$)  &  ($M_{\odot}$)  &
	 & ($10^{-15}$ g cm$^{-3}$)  &    &  ($10^4~R_{\odot}$)  & ($10^{51}$ erg)  \\

\hline
sn94w43 &  12                 &  8      &  0.015        &
  0.319  &    5     &   2         &  12  & 1.07 \\

sn94w58 &  7            &      2 &  0.015        &
 0.31    &    12     &      1         &    6.6  & 1.25 \\

sn94w64 &  7            &      2 &  0.015        &
 0.31    &    0     &      -         &    -  & 1.33 \\
\hline
\end{tabular}
\end{minipage}
\label{hydro}
\end{table*}



Each model was exploded by the deposition of heat energy in a layer of
mass $\sim 0.06~ M_\odot$ outside of $1.41~ M_\odot$. Since {\sc stella}
does not include nuclear burning, preservation of the same mixed
composition in the ejecta is assured.

We explored parameter space for the mass ($M$) and energy ($E$) of
the SN ejecta and found acceptable fits to the data for masses in the
range 6--15~$M_{\odot}$ and with the energy given by the ratio
$E/M \approx 0.15$--0.2 foe~$M^{-1}_{\odot}$.
In the following sections and in Table \ref{hydro}, we present three
representative models which allow us to demonstrate different aspects
of the formation of the light curve.  Of these three, model sn94w58 provides
our best fit to SN 1994W, model sn94w43 shows that a normal SN II-P light
curve is not appropriate for SN 1994W, and model sn94w68 illustrates what
the supernova light curve might have looked like without CS
interaction.  All three models have $E = 1.5 \times 10^{51}$ ergs.
This energy is typical for good light-curve models.  The asymptotic
kinetic energy of the ejecta is somewhat lower and is given in the
Table \ref{hydro}.

\subsection{Hydrodynamics and shock propagation}

\begin{figure}
  \centering
  \vspace{8cm}
  \includegraphics{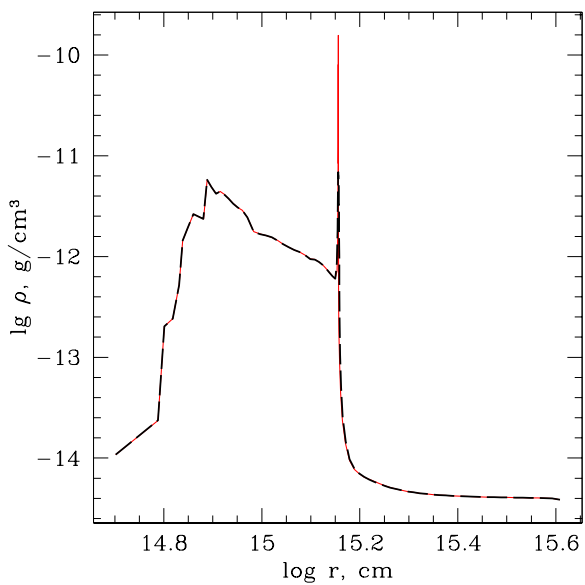}
  \caption[]{Density as a function
of radius ($r$) in two models with different amounts of artificial
smearing at day 30. The height of the density spike with small smearing
({\it solid line}) reaches the theoretical estimate for an isothermal shock wave.
  }
  \label{rhor2}
  \end{figure}

The modelled light curves are dominated by the diffusion of the
trapped radiation generated during the shock wave's propagation in the
extended stellar atmosphere and subsequent emission of a radiative
shock propagating in the dense circumstellar medium.  A  dense
shell is formed which is found in non-equilibrium radiation
hydrodynamic modelling but often missed in equilibrium diffusion
modelling.  We identify this dense shell as the CDS discussed in
Section \ref{sec-cds}. The Lagrangian code {\sc stella} does a good job of
resolving the very fine structure of the opaque shell, which contains
approximately one solar mass of the material (Fig.\ \ref{rhor2}).

\subsection{Light curves}

\begin{figure}
  \centering
  \vspace{8cm}
  \includegraphics{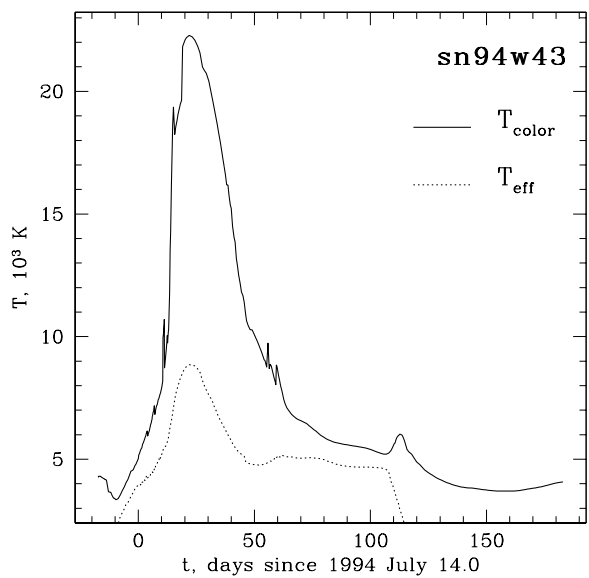}
  \caption[]{Colour and effective
temperatures for the model sn94w43. Realistic, scattering-dominated opacity
has been assumed. The solid line shows the temperature of the best black-body fit
to the flux (colour temperature). The dashed line shows the effective
temperature defined by the luminosity and the radius of last scattering.}
  \label{tphea43}
  \end{figure}

\begin{figure}
  \centering
  \vspace{8cm}
  \includegraphics{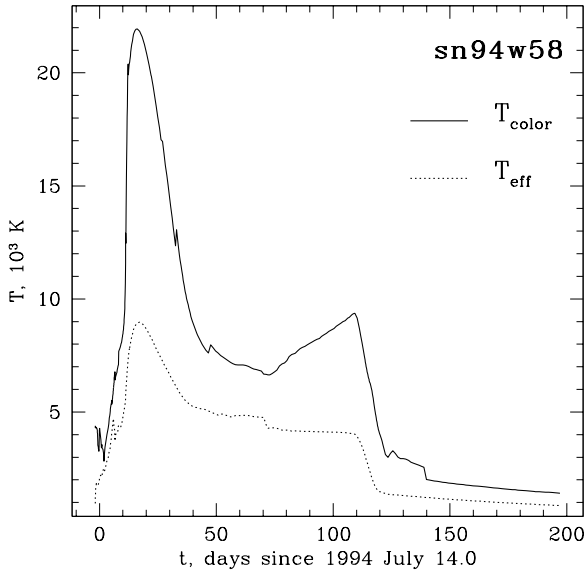}
  \caption[]{Colour and effective temperatures for the model sn94w58.}
  \label{tphea58}
  \end{figure}

In Figs.\ \ref{tphea43} and \ref{tphea58} we show the changes in the
colour temperature, $T_{\rm c}$, of the best black-body fit to the
flux.  We compare this to the effective temperature, $T_{\rm eff}$,
defined by the luminosity and the radius of last scattering $R$
through $L=4\pi\sigma T^4_{\rm eff}R^2$ (see \citealt{blin93j} for
details of finding $R$ and from that $T_{\rm eff}$).  Our multi-group
radiative transfer with hydrodynamics obtains this temperature in a
self-consistent way, and no additional estimates of the thermalization
depth are needed (in contrast to the one-group model of
\citealt{ens92}, for example).  The large difference between colour
and effective temperatures is partly due to a geometric effect (the
radius of last scattering is greater than the effective radius of
photon creation) and partly due to the blanketing effect of scattering
(the average energy of the photons is higher than that corresponding
to the value of $T_{\rm eff}$; \citealt{schuster05}).  The modelled
$T_{\rm c}$ changes in a way similar to the values derived from
black-body fits to the spectra (Section \ref{sec-continuum}).

\begin{figure}
  \centering
  \vspace{8cm}
  \includegraphics{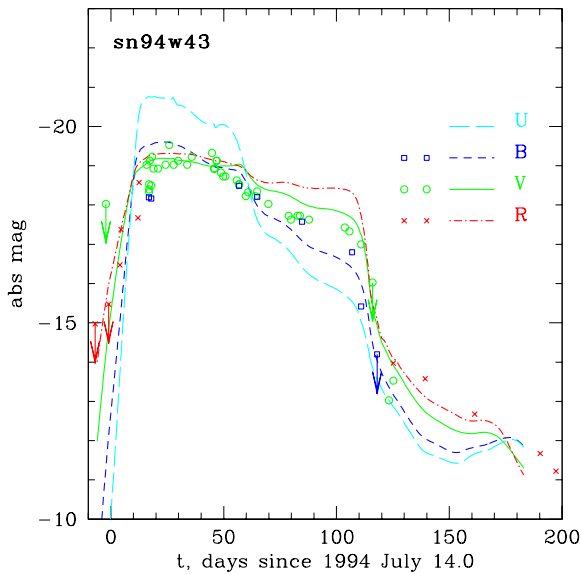}
  \caption[]{Light curves for the model sn94w43. The photometric points
  are taken from SCL98 and references therein.}
  \label{lc43}
  \end{figure}

\begin{figure}
  \centering
  \vspace{8cm}
  \includegraphics{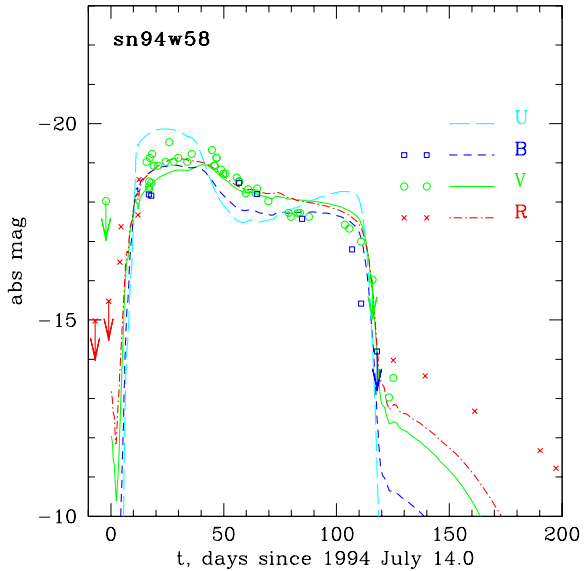}
  \caption[]{Same as in  Fig.~\protect\ref{lc43}, but for model sn94w58. }
  \label{lc58}
  \end{figure}

\begin{figure}
  \centering
  \vspace{8cm}
  \includegraphics{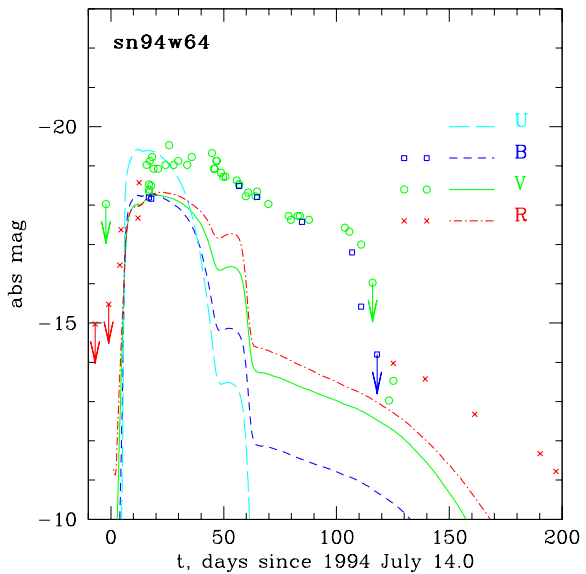}
  \caption[]{Same as in  Fig.~\protect\ref{lc43}, but for model sn94w64. }
  \label{lc64}
  \end{figure}

The light curves are shown in Figs.~\ref{lc43}, \ref{lc58} and
\ref{lc64}.  The model sn94w43 produces a bright light curve, but the
second half of its plateau is dominated by diffusion, not by the
shock, so the behaviour of colours is similar to that of a typical
SN~II-P.

The density of the more powerful wind in the model sn94w58 follows the
law $\rho \propto r^{-1}$.
This model produces a much better fit to the colours at the late
plateau phase.  The flat CS density distribution is consistent with
the modelling of H$\alpha$ evolution (Section \ref{sec-haev}). Model
sn94w64, which is exactly the same as model sn94w58 in the bulk mass,
but does not have the powerful wind, is much less luminous at the late
plateau phase. This shows the importance of CS interaction for the
luminosity.

\begin{figure}
  \centering
  \vspace{8cm}
  \includegraphics{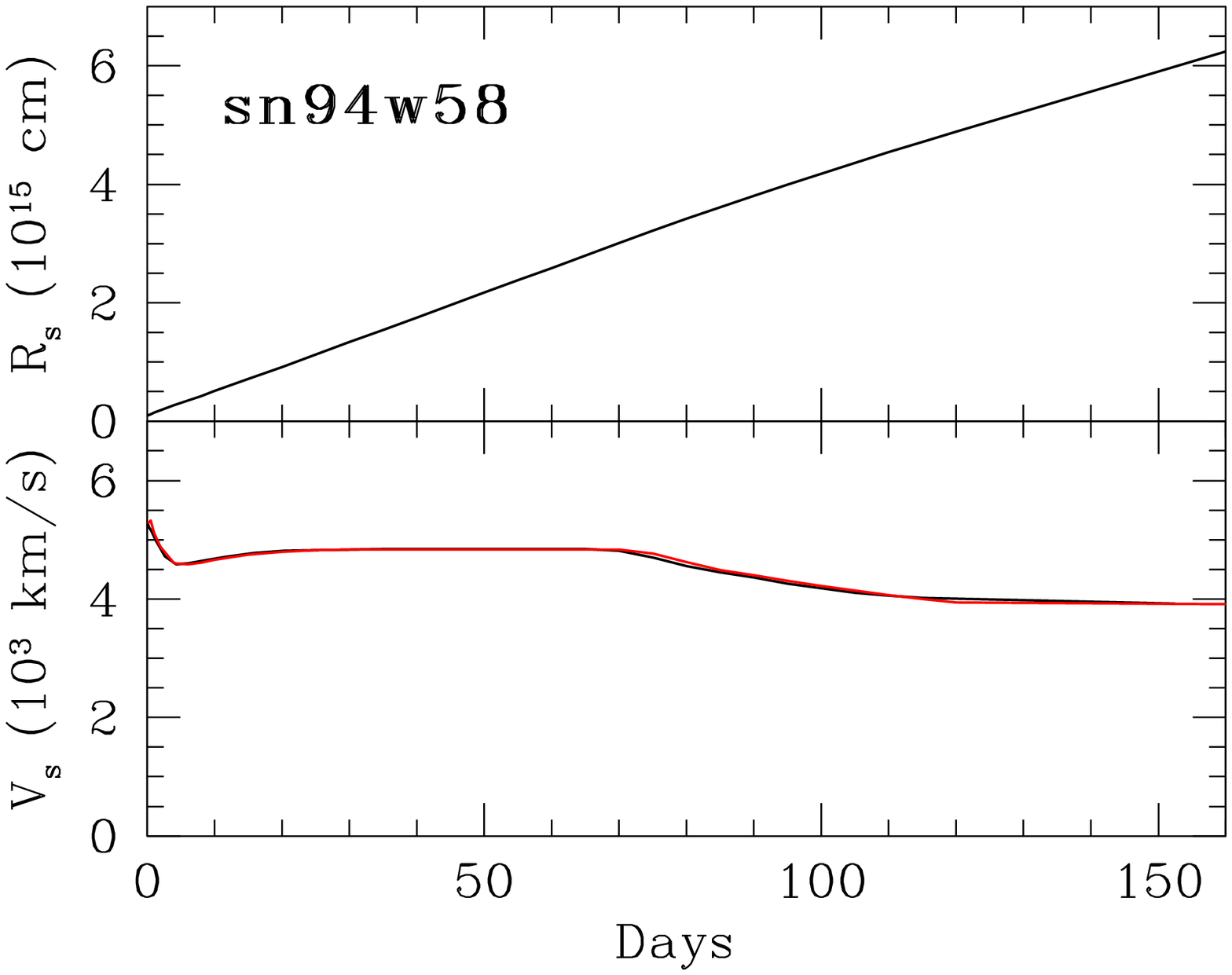}
  \caption[]{The evolution of the radius and velocity of the cool
  dense shell in hydrodynamical model sn94w58.
  Compare with Fig.~\protect\ref{f-dyn}.
  }
  \label{rvsh}
  \end{figure}

The time-dependence of the radius $R_s$ and velocity $V_s$ of the cool dense
shell in model sn94w58 is shown in  Fig.\ \ref{rvsh}.
The two nearly indistinguishable lines in the $V_s$ plot are obtained by
taking $V_s=dR_s/dt$ numerically, and by taking the mass-averaged
speed of matter inside the shell. A comparison with Fig.~\ref{f-dyn}
shows good agreement with the spectral model, although $V_s$ is somewhat
higher in the hydrodynamical model. The difference is at most $\sim
10$ per cent.

To summarize, the hydrodynamical modelling of the light curve
of SN~1994W suggests that the optimal light curve is produced in
a model which contains a dense CS envelope at $r \gsim 10^{15}$ cm
with a relatively flat density distribution and an outer radius
of $\sim 4.5 \times 10^{15}$ cm.  This conclusion is fully
consistent with the H$\alpha$ modelling.

\section{Discussion and Conclusions} \label{sec-discussion}

We have presented and analysed spectra and light curves of SN~1994W,
one of the best-observed SNe~IIn.  During the first three months the
spectrum clearly shows the presence of three line components: (i)
narrow P-Cygni lines, (ii) intrinsically broad lines and (iii)
extended smooth wings in H$\alpha$ and H$\beta$.  We attribute these
components, respectively, to (i) a dense CS envelope, (ii) shocked,
cool, dense gas confined in a narrow layer on top of the photosphere,
and (iii) the effects of Thomson scattering in the CS envelope.

Our line profile analysis and hydrodynamical light-curve modelling
have led us to a coherent picture.  SN~1994W appears to have been the
result of explosion of an extended pre-supernova ($\sim 10^{15}$ cm)
embedded in an extended CS envelope [$\sim (4-5) \times 10^{15}$ cm]
with Thomson optical depth $\sim 2.8$ one month after the explosion.

Although we do not rule out the presence of a wind from a normal red supergiant
pre-supernova outside the dense CS envelope, such a wind component
cannot have been very dense, considering the low luminosity of
H$\alpha$ on day 203 ($L\approx 10^{38}$ erg s$^{-1}$).  This luminosity
is two orders of magnitude lower than for SN~1979C ($\sim10^{40}$ erg
s$^{-1}$) at $\sim 1$ yr \citep{cf85}.  Attributing this difference to
the difference in the interaction luminosity ($L\propto 4\pi\rho
v_{\rm s}^3$), we can estimate the density parameter of the outer wind
in SN~1994W. The expansion velocity of the CDS in SN~1994W, $\sim
4000$ km s$^{-1}$, is half that of SN~1979C ($\sim 8000$ km s$^{-1}$).
Given the SN~1979C wind parameter, $w \approx 10^{16}$ g cm$^{-1}$, the
outer wind in SN~1994W should have $w \approx 10^{15}$ g cm$^{-1}$ to
account for the difference in the H$\alpha$ luminosity.  This is at
least ten times lower than in SN~1979C, and a factor of two lower than
in SN~1980K \citep[e.g., ][]{lf88}.  In this respect, the detection by
\citet{schlegel99} of X-rays from
SN~1994W at $t = 1180$ days with luminosity $L \approx 8 \times 10^{39}$ erg
s$^{-1}$ appears odd when compared with the expected X-ray luminosity
($\lsim10^{38}$ erg s$^{-1}$) at that epoch for a case of SN/wind
interaction with $w \approx 10^{15}$ g cm$^{-1}$.  To resolve this problem,
one would need the wind density to be at least of factor of ten higher
at $\sim 4\times10^{16}$ cm, i.e., exceeding that of SN~1979C.  This
could argue for a possible multi-shell ejection scenario prior to the
explosion.

The recovered kinematics, density and linear scale of the CS matter
around SN 1994W imply a kinematic age for the CS envelope of $t_{\rm
cs}\approx 1.5$ yr, mass $M_{\rm cs} \approx 0.4~M_{\odot}$ and kinetic
energy $E_{\rm cs}\approx2\times10^{48}$ erg.  The enormous average
mass-loss rate ($M_{\rm cs}/t_{\rm cs} \approx 0.3~M_{\odot}$ yr$^{-1}$)
and equally enormous kinetic luminosity ($E_{\rm cs}/t_{\rm
cs} \approx 4 \times 10^{40}$ erg s$^{-1}$) of the mass-ejection mechanism
strongly suggest that the CS envelope has been lost as a result of an
explosive event which occurred $\sim 1.5$ yr prior to the SN
outburst.

In their model for the narrow lines in SN~1983K, \citet{grasberg86}
invoked the violent ejection of the hydrogen envelope 1--2 months
before the SN explosion with a velocity on the order of $\sim 2500$
km s$^{-1}$ and an energy $(1-2) \times 10^{49}$ erg.
Dramatically high mass loss shortly before explosion has also been derived
for the SN~IIn 1995G by \citet{cd03}, on the basis of an analysis
similar to the one we have presented here.

The mechanism behind such violent ejections might be associated with a
nuclear flash in the degenerate core of the pre-supernova.  This
conjecture, also presented by \citet{cd03}, is prompted by the
prediction of \citet{ww79} that the O/Ne/Mg core of an $\sim
11~M_{\odot}$ star may produce strong Ne flashes several years prior
to the SN explosion and that the strongest flash could eject most of the
hydrogen envelope with velocities of $\sim 100$ km s$^{-1}$.  Although
such flashes do not occur in more recent models with finer zoning
(\citealt*{woosley02}), we believe that the effect warrants further
investigation, especially given the highly complicated nuclear burning
regime and hydrodynamics
of degenerate O/Ne/Mg cores.

Remarkably, the low $^{56}$Ni mass
($<0.015~M_{\odot}$) estimated from the tail $R$-band luminosity of
SN~1994W (SCL98) seems to be in accord with the suggested mass of such
a progenitor, if at the lower end of our estimated mass range.  Such
stars are expected to eject only small amounts of $^{56}$Ni
\citep{mw88}.  Our models are, however, not very sensitive to the zero-age
main-sequence mass of the progenitor, and, as SCL98 point out, the low
nickel content could indicate either a low-mass progenitor which may have
lost a few solar masses in shell and wind ejections, or a more massive
star that lost several solar masses prior to the ejection of the CS
shell.  Deep X-ray searches for more extended CS material could help solve this
problem, as should more detailed models for stellar evolution just
prior to core collapse.

How common are events like SN~1994W?  Of the SNe~IIn that have been
observed so far, it is possible that a close counterpart has been seen
but not recognised as such due to sparser temporal coverage or greater
distance.  Nevertheless, several objects
have shown narrow hydrogen and \feii\ absorption lines with
velocities of about 1000 km s$^{-1}$.  This family includes
SNe~1987B \citep{schlegel96}, 1994aj
\citep{benetti98}, 1994ak \citep{filippenko97}, 1995G
\citep{pastorello02}, 1996L
\citep{benetti99} and 1999el \citep{dicarlo02}.  We suggest that  in
at least some of these SNe, a CS envelope was ejected in a
violent, explosive manner, as we believe was the case for SN~1994W,
and as \citet{cd03} have suggested for SN 1995G.  Further data and
analyses are needed, however, before it can be demonstrated that the
energy of the ejected CS envelope is comparable for these events.

In any case, we suggest that at least part of the variety of SNe~IIn
has been accounted for.  Some, like SN~1988Z and SN~1998S, have a
dense, slow CS envelope formed by a slow superwind.  Others, like SN
1994W, seem to have ejected a CS envelope in a violent event a few
years before the SN explosion.

\section*{Acknowledgements}

We thank Aaron J. Barth, Mike Breare, Ren{\'e} Rutten, Luis C. Ho,
Neil O'Mahony, Chien Y.\ Peng and Ed Zuiderwijk for helping take
observations for us, and Clive Jackman, Danny Lennon and Marco Azzaro
for help assessing the day 31 spectrum calibration.  We also thank
Itziar Aretxaga, Eddie Baron, Claes Fransson, Seppo Mattila, Peter
Meikle, Miguel P{\'e}rez Torres, and Luca Zampieri for discussions.

This paper is based on observations made with the Isaac Newton
Telescope (INT), the William Herschel Telescope (WHT), and the Nordic
Optical Telescope (NOT).  The INT and WHT are operated on the island
of La Palma by the Isaac Newton Group in the Spanish Observatorio del
Roque de los Muchachos of the Instituto de Astrof{\'\i}sica de
Canarias. The NOT is operated on the island of La Palma jointly by
Denmark, Finland, Iceland, Norway, and Sweden, in the Spanish
Observatorio del Roque de los Muchachos of the Instituto de
Astrof{\'\i}sica de Canarias.  Some of the data presented herein were
obtained at the W.M. Keck Observatory, which is operated as a
scientific partnership among the California Institute of Technology,
the University of California and the National Aeronautics and Space
Administration. The Observatory was made possible by the generous
financial support of the W.M. Keck Foundation.

This project was supported by the Royal Swedish Academy of Sciences.
S.I.B.\ was supported partly by RFBR 02-02-16500, the Wenner-Gren
Science Foundation, MPA Garching and ILE Osaka guest programs. The
research of P.L.\ is sponsored by the Royal Swedish Academy and the
Swedish Research Council, and he is a Research Fellow at the Royal
Swedish Academy supported by a grant from the Wallenberg
Foundation. A.V.F.'s research is supported by National Science
Foundation grant AST-0307894.


\bsp

\label{lastpage}

\end{document}